\documentclass[aps,prb,longbibliography,twocolumn,floatfix,superscriptaddress,nofootinbib]{revtex4-2}

\usepackage{graphicx}
\usepackage{amsmath}
\usepackage{amssymb}
\usepackage{mathtools}
\usepackage{makeidx}
\usepackage{amsfonts}
\usepackage{bm}
\usepackage{tikz}
\usepackage{xcolor} 

\usepackage[colorlinks=true, linkcolor=blue, citecolor=blue, urlcolor=blue]{hyperref}
\usetikzlibrary{shapes,shadows}
\newcommand{\ket}[1]{\lvert #1 \rangle}

\begin{document}

\title{Floquet mobility edges and transport in a periodically driven generalized Aubry-Andr\'e model}

\author{Jayashis Das}
\email{jayashis24@iiserb.ac.in}
\affiliation{Department of Physics, Indian Institute of Science Education and Research, Bhopal, Madhya Pradesh 462066, India}
\author{Vatsana Tiwari}
\affiliation{Department of Physics, Indian Institute of Technology Bombay, Mumbai 400076, India}
\author{Manish Kumar}
\affiliation{Department of Physics, Indian Institute of Science Education and Research, Bhopal, Madhya Pradesh 462066, India}
\author{Auditya Sharma}
\email{auditya@iiserb.ac.in}
\affiliation{Department of Physics, Indian Institute of Science Education and Research, Bhopal, Madhya Pradesh 462066, India}
\date{\today}

\begin{abstract}

We investigate the effect of a periodic electric field drive on the
generalized Aubry-Andr\'e model, also known as the Ganeshan-Pixley-Das
Sarma (GPD) model, which is well known as a host of mobility
edges. Our study of the Floquet spectrum of the driven GPD model
uncovers the emergence of two distinct Floquet mobility edges, a
delocalized--localized (DL) edge in the bounded regime, and a
multifractal--localized (ML) edge in the unbounded regime. By
  mapping the high-frequency effective Hamiltonian to the static GPD
  model via a Bessel-renormalized hopping, the driven mobility edge
  formulae follow directly, with drive-induced localization arising
  naturally at specific values of the driving parameters, where the
  effective tunneling is coherently suppressed. Combined with
numerical diagnostics such as the fractal dimension and inverse
participation ratio, we demonstrate that these mobility edges can be
effectively controlled by the amplitude and frequency of the electric
field drive. Furthermore, the dynamical study of the periodically
driven GPD model demonstrates superdiffusive to almost ballistic
transport in the bounded regime corresponding to the DL edges, whereas
subdiffusive transport is observed in the unbounded regime associated
with the ML edges. We also analyze deviations from the high-frequency
effective description by explicitly examining the low-frequency
driving regime, where significant and counterintuitive deviations in
both spectral properties and transport behavior are observed. Our
study highlights the interplay of a quasiperiodic potential and a
periodically varying electric field drive as a powerful mechanism to
engineer mobility edges and control transport in systems with rich
spectral features.
\end{abstract}

\maketitle
\section{Introduction}
\label{sec:introduction}

Understanding localization and transport in quantum systems remains a fundamental problem in condensed matter physics. A cornerstone of this field is Anderson localization~\cite{Anderson_localization}, which shows that even weak onsite disorder can localize wavefunctions and suppress transport. In one-dimensional systems with random disorder, all the single particle eigenstates are typically localized~\cite{AL_2D,2DAnderson}, whereas in higher dimensions mobility edges can emerge~\cite{3D_ME,3D_ME2}, separating localized and extended states. 
Quasiperiodic potentials provide an alternative route to delocalization--localization transitions even in one dimension. A well-known example is the one-dimensional Aubry-Andr\'e-Harper (AAH)~\cite{aubry,harper1955single} model, which exhibits a self-dual transition between localized and extended phases, but lacks mobility edges. Generalizations that break this self-duality via long-range hopping or modified and diluted potentials~\cite{GAAMOBILITYEDGE,GAAH,Mosaic,ME_TAPAN,LRH,PhysRevLett.123.070405, 10.21468/SciPostPhys.12.1.027, EXACT_LONG_RANGE_ME,SV_ME,Gen_GAA,GEN_AAH, PhysRevB.106.054204} lead to energy-dependent localization transitions, giving rise to mobility edges even in one dimension. These phenomena have been explored in a variety of experimental platforms like photonic lattices~\cite{GAO202558,PhysRevResearch.5.033170,PhysRevLett.103.013901}, ultracold atoms~\cite{PhysRevA.92.043606,PhysRevLett.120.160404,PhysRevLett.126.040603, sunil2026localizationhoppingdisorderquasiperiodic}, and Rydberg atom
arrays~\cite{yoshida2024proposalexperimentalrealizationquantum}. The generalized Aubry-Andr\'e (GAA) model, also known as Ganeshan-Pixley-Das Sarma (GPD) model~\cite{GAAH}, is a paradigmatic example where a modification in the quasiperiodic potential breaks the self-duality, leading to the emergence of mobility edges. In particular, the spectrum hosts two distinct types of transitions, delocalized--localized (DL) and multifractal--localized (ML) edges~\cite{GAAMOBILITYEDGE}, upon varying the quasiperiodic onsite potential.
Its analytically tractable mobility edges and coexistence of distinct phases make it an ideal platform to investigate energy-resolved localization and multifractality, and thus form the basis of the present work.

Recently, increasing attention has been devoted to understanding how periodic driving influences localization phenomena in single particle~\cite{dyn_loc1,dyn_loc2,Devendra_dyn,Vatsana_dyn,periodically2023aditya} as well as many-body quantum systems~\cite{Moessner_MBL,Ponte_MBL,driven_stark_MBL}. Within Floquet theory~\cite{Floquet1,Floquet2}, the long-time evolution of a periodically driven system is governed by an effective Floquet Hamiltonian~\cite{Eckardt_2015,High_frequency_H2,High_frequency_H3}, whose parameters depend on the drive amplitude and frequency. Time-periodic modulation thus provides a powerful means of dynamically tuning quantum systems, enabling a wide range of phenomena, including dynamical localization~\cite{dunlap1988dynamic,dyn_loc2,Driven_MBL}, coherent destruction of tunneling~\cite{Coherent1,Coherent2,Coherent3}, Floquet-engineered topological phases~\cite{Floquet_Topological,Topological_2,Topological_3,Topological_4,Topological_5,Topological_6}, and control of quantum phase transitions~\cite{QPT_1,QPT_2}, with applications in ultracold atoms and photonic lattices~\cite{Goldman2016,Optical_lattice_2,Optical_lattice_2,Optical_lattice_3,Optical_lattice_4,WELDPRR}. In quasiperiodic systems, periodic driving can further modify localization properties and induce mobility edges~\cite{Madhumita_ME,Driven_ME,PhysRevB.111.174202} that depend on driving parameters. While Floquet engineering of mobility edges is demonstrated in the mosaic AAH model~\cite{FloquetME}, that work is restricted to spectral properties of delocalized--localized (DL) edges in the high-frequency limit. In contrast, the driven GPD model studied here hosts both DL and multifractal--localized (ML) Floquet mobility edges, and we further investigate transport and quantum dynamics across these distinct regimes~\cite{Moessner_MBL,Madhumita_ME,Floquet_Topological,periodically2023aditya}. Furthermore, we explore low-frequency driving regimes to demonstrate how higher-order Floquet processes modify both spectral properties and transport beyond the high-frequency limit.

In this work, we study the periodically driven GPD model subjected to a sinusoidal electric field drive. Using the Floquet–Magnus expansion in the high-frequency regime, we derive an effective static Hamiltonian where the hopping amplitude is renormalized by a zeroth-order Bessel function whose argument depends on the driving parameters. This framework allows us to extend the static GPD results to the driven case by mapping the mobility edges obtained from Avila's global theory~\cite{Avila2015, GAAMOBILITYEDGE} onto the Floquet spectrum. Here, drive-induced localization arises naturally at the zeros of the Bessel function as a result of coherent suppression of tunneling\cite{Dynamical_localization,dyn_loc1,dyn_loc2}. These analytical predictions show excellent agreement with numerical diagnostics, including the fractal dimension, inverse participation ratio, and standard deviation of Floquet eigenstates. Beyond spectral properties, we explore the dynamical consequences of these mobility edges using multiple transport probes, including the root-mean-squared deviation of wave packets, the growth of entanglement entropy, and the return probability. These observables reveal distinct transport regimes associated with drive induced localization and mobility edges. Finally, we investigate deviations from the high-frequency limit, uncovering nontrivial spectral and dynamical behavior at lower driving frequencies. Overall, our findings demonstrate that periodically driven quasiperiodic systems offer a powerful platform for engineering and probing mobility edges and quantum transport phenomena.

The paper is organized as follows. In Sec.~\ref{sec:model}, we introduce the model and discuss its static limit. Section~\ref{sec:floquet_ME} presents the derivation of the effective Floquet Hamiltonian and the analytical mobility edges. In Sec.~\ref{sec:numerics}, we present numerical results based on the fractal dimension, inverse participation ratio, and standard deviation of eigenstates, and examine the effects of low-frequency driving. The real-time dynamics and transport properties are then explored in Sec.~\ref{sec:Dynamics}. Finally, we summarize our findings and conclude in Sec.~\ref{sec:conclusion}.

\section{Model}
\label{sec:model}
We investigate the one-dimensional periodically driven Ganeshan-Pixley-Das Sarma (GPD) model~\cite{GAAH, PhysRevLett.126.040603}, describing non-interacting spinless fermions on a quasiperiodic lattice. The time-dependent Hamiltonian for this system is given by
\begin{equation}
\label{hamiltonian}
\hat{H}(t) =
\sum_{j=1}^{L-1}
\Big(
J\, \hat{c}_{j}^{\dagger} \hat{c}_{j+1}
+ \text{H.c.}
\Big)
+ \sum_{j=1}^{L}
\big[V_j + A_j(t)\big]\,
\hat{c}_{j}^{\dagger} \hat{c}_{j},
\end{equation}
where $\hat{c}_j^{\dagger}$ ($\hat{c}_j$) creates (annihilates) a fermion at site $j$, and $L$ denotes the total number of lattice sites. 
The quasiperiodic onsite potential is given by
\begin{equation}
\label{quasiperiodic_potential}
V_j =
\frac{2\lambda \cos(2\pi \alpha j + \phi)}
{1 - \beta \cos(2\pi \alpha j + \phi)},
\end{equation}
where $\lambda$ denotes the quasiperiodic potential strength and $\phi \in (0,2\pi)$ is an arbitrary global phase. The irrational number $\alpha$ controls the incommensurability. We choose $\alpha = (\sqrt{5}-1)/2$, the inverse of the golden mean, which is standard in numerical studies~\cite{goldenmean,LRH}. Unless otherwise stated, we fix the system size to be $L=987$, set $J=1$, and take the quasiperiodic potential strength to be $\lambda=0.05$. The parameter $\beta$ deforms the potential, qualitatively changing its spectral properties.  The time-periodic electric field drive is given by 
\begin{equation}
\label{drive}
A_j(t) = - j K \cos(\omega t),
\end{equation}
with driving amplitude $K$ and frequency $\omega$. Such driving protocols are routinely realized in ultracold-atom
experiments via lattice shaking~\cite{lattice_shaking} of optical potentials~\cite{WELDPRR,WELDPRX},
providing a highly controllable platform for Floquet engineering
of quasiperiodic systems.
In the static limit ($K=0$), for $\beta=0$, the model described in Eq.~\eqref{hamiltonian} reduces to the AAH model~\cite{aubry,harper1955single}, which exhibits delocalized single particle eigenstates for $\lambda < J$ and exponentially localized states for $\lambda > J$, while at the self-dual point $\lambda = J$, all eigenstates are multifractal. For $|\beta|<1$, the potential is bounded, thereby making the Hamiltonian a bounded operator and consequently, all the eigenvalues of $\hat{H}$ are finite. On the other hand for $|\beta|\geqslant1$, the potential becomes unbounded, leading to a set of eigenvalues of $\hat{H}$ that are unbounded~\cite{GAAMOBILITYEDGE,GAAH}.
\subsection{Mobility edges in the undriven model}
We first summarize the properties of the undriven model, corresponding to the static GPD Hamiltonian. Unlike the standard AAH model, the GPD model hosts exact single particle mobility edges separating eigenstates with distinct localization properties. For $|\beta|<1$, extended and localized eigenstates coexist in energy and are separated by a DL edge~\cite{GAAMOBILITYEDGE} given by
\begin{equation}
\label{eq:static_ME_bounded}
\left|\frac{\beta E_c + 2\lambda}{J}\right| = 2 \qquad (\text{for } |\beta|<1),
\end{equation}
which determines the critical energy $E_c$ separating delocalised and localized states. The position of the mobility edge is independent of the phase $\phi$ and system size.
For $|\beta|\geqslant1$, the nature of the transition changes qualitatively, resulting in a ML edge. The exact mobility edge condition in this regime becomes~\cite{GAAMOBILITYEDGE}
\begin{equation}
\label{eq:static_ME_unbounded}
\left|\frac{\beta E_c + 2\lambda}{J}\right| = 2|\beta| \qquad (\text{for }|\beta|\geqslant1).
\end{equation}
\begin{figure}
    \centering
    \includegraphics[width=0.48\textwidth]{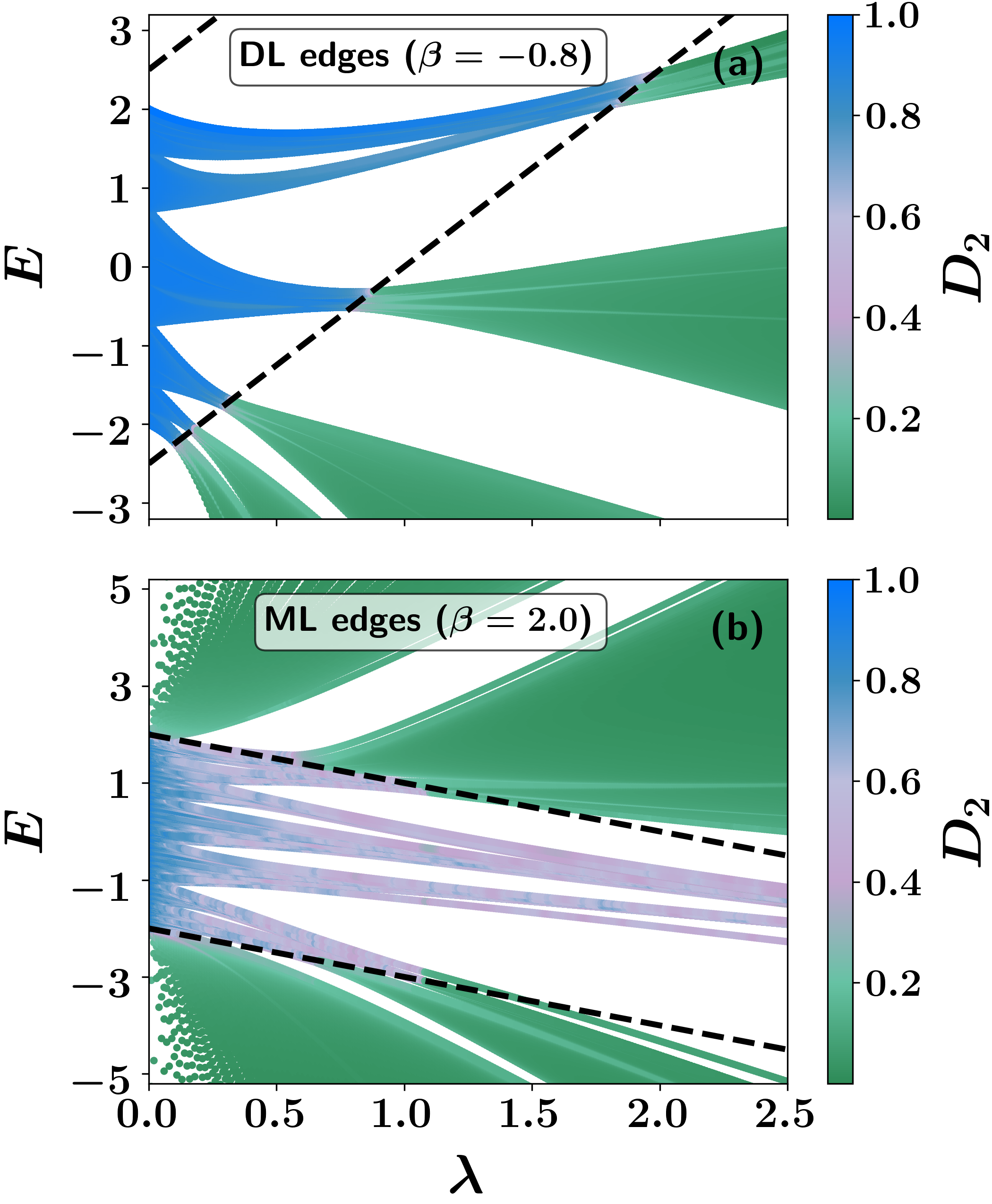}

    \caption{
Fractal dimension $D_2$ of all the eigenstates of the undriven GPD model as a function of eigenenergies $E$ and quasiperiodic potential strength $\lambda$. (a) For $\beta=-0.8$, where extended and localized eigenstates coexist and are separated by a DL edge. (b) For $\beta=2.0$, where the extended phase is replaced by a multifractal phase, leading to an ML edge. 
Black dashed lines denote the exact analytical mobility edges given in Eq.~\eqref{eq:static_ME_bounded} and \eqref{eq:static_ME_unbounded}. Results are shown for a system size $L = 987$ with periodic boundary conditions, averaged over 10 realizations of the phase $\phi$.}
    \label{fig:static_ME}
\end{figure}

To characterize these phases numerically, we compute the fractal dimension $D_2$ of single particle eigenstates of the Hamiltonian in Eq.~\eqref{hamiltonian}, with $K=0$. The fractal dimension $D_2$ of an eigenstate is given by~\cite{D2_Fit},
\begin{equation}
\label{fractal_dimension}
D_2 =
-\lim_{L\to\infty}
\frac{\ln \!\left(\sum_{j=1}^{L} |\psi_{n,j}|^4\right)}
{\ln L}
\end{equation}
where $\psi_{n,j}$ is the amplitude of the $n^{\text{th}}$ normalized eigenstate at site $j$. In the thermodynamic limit, $D_2=1$ corresponds to extended states, $D_2=0$ to exponentially localized states, and $0<D_2<1$ indicates multifractal (critical) states. Fig.~\ref{fig:static_ME} shows the plot of $D_2$ of all the eigenstates as a function of $\lambda$ and eigenenergy $E$. For $\beta=-0.8$, the spectrum exhibits a DL edge [Fig.~\ref{fig:static_ME}(a)]. For $\beta=2.0$, the extended phase is replaced by a multifractal phase, yielding an ML edge [Fig.~\ref{fig:static_ME}(b)]. In both panels, the black dashed curves represent the exact mobility edge conditions given by Eqs.~(\ref{eq:static_ME_bounded}) and (\ref{eq:static_ME_unbounded}).

In the next section, we examine the periodically driven model, showing how the high-frequency Floquet expansion allows us to directly map these static mobility edge conditions to the driven regime via an effective Bessel-renormalized hopping amplitude.
\section{High-frequency Floquet Hamiltonian and analytical mobility edges}
\label{sec:floquet_ME}

\subsection{High-frequency effective Floquet Hamiltonian}
We now turn to the periodically driven case ($K\neq 0$), where the system is governed by the time-periodic Hamiltonian given in Eq.~\eqref{hamiltonian}. The linear drive in Eq.~\eqref{drive}, corresponding to a time-dependent electric field, admits a natural gauge interpretation and can be treated analytically by transforming to a rotating frame. To eliminate the explicit time dependence associated with the drive, the first step is to introduce a unitary transformation~\cite{Bukov04032015,FloquetME}
\begin{equation}
\label{unitary_transformation}
\hat{R}(t)
=
\exp\!\left[
i \sum_{j} f_j(t)\, \hat{c}_j^\dagger \hat{c}_j
\right],
\end{equation}
with the generator
\begin{equation}
\label{unitary_generator_function}
f_j(t)
=
-\int_{t_0}^{t} A_j(t')\,dt'
- j\bar f
=
jF(t).
\end{equation}
The constant $\bar f$ is chosen such that $\int_0^T f_j(t)\,dt=0$, ensuring that $\hat{R}(t)$ is strictly periodic over one driving period $T=2\pi/\omega$. For the cosine drive considered here, one finds $F(t)=\frac{K}{\omega}\sin(\omega t).$ Under this transformation, the Hamiltonian in the rotating frame is given exactly by
\begin{equation}
\hat{H}_R(t)
=
\hat{R}^\dagger(t)\hat{H}(t)\hat{R}(t)
-
i\hat{R}^\dagger(t)\frac{d}{dt}\hat{R}(t),
\end{equation}
which evaluates to
\begin{equation}
\label{rotating_frame_Hamiltonian}
\hat{H}_R(t)
=
\sum_j
\Big[
J e^{iF(t)} \hat{c}_j^\dagger \hat{c}_{j+1}
+
J e^{-iF(t)} \hat{c}_{j+1}^\dagger \hat{c}_j
\Big]
+
\sum_j V_j \hat{c}_j^\dagger \hat{c}_j .
\end{equation}
Thus, the periodic drive manifests itself solely through a
time-periodic Peierls phase in the hopping term, while the
quasiperiodic onsite potential remains unchanged. Importantly, this
step involves no approximation and holds for arbitrary driving
strength and frequency. The dynamics of the driven system is most
naturally analyzed within Floquet theory ~\cite{Floquet1,Floquet2},
where the stroboscopic evolution over one driving period is generated
by an effective time-independent Floquet Hamiltonian $\hat{H}_F$, such
that the time-evolution operator over one period (Floquet operator) is
given by
\begin{equation}
\hat{U}(T) = \mathcal{T}\exp\left(-i \int_{0}^{T} \hat{H}(t)\,dt \right) = \exp(-i \hat{H}_F T).
\label{Floquet_operator}
\end{equation}
In general, $\hat{H}_F$ is difficult to compute exactly due to time
ordering. However, when the driving frequency $\omega$ is the dominant
energy scale of the problem, i.e. $\omega \gg J$ and $\omega \gg
\lambda$ ($\hbar$ is set to $1$ overall), $\hat{H}_F$ can be
systematically approximated using a high-frequency (Floquet--Magnus)
expansion~\cite{Eckardt_2015,Bukov04032015,TAKASHI_MORI,High_frequency_H2,High_frequency_H3,Sen_2021}. The
effective Floquet Hamiltonian is given by $\hat{H}_F = \hat{H}_F^{(0)}
+ \hat{H}_F^{(1)} + \hat{H}_F^{(2)} + \cdots$, where $\hat{H}_F^{(n)}$
scales as $\omega^{-n}$. To leading (zeroth) order,
$\hat{H}_F\simeq\hat{H}_F^{(0)}$ (in the high-frequency limit) is
given by the time average of the rotating-frame Hamiltonian
\begin{equation}
\label{High_frequency_approximated_H}
\hat{H}_F^{(0)}
=
\frac{1}{T}
\int_0^T \hat{H}_R(t)\,dt.
\end{equation}
Evaluating the time average using the Jacobi--Anger expansion~\cite{Dar_2025}, $e^{ix\sin\theta}=\sum_{\ell=-\infty}^{\infty}J_\ell(x)e^{i\ell\theta}$, we find that all oscillatory contributions vanish upon averaging, leaving only the $\ell=0$ component. This yields the effective Floquet Hamiltonian in the leading order as
\begin{equation}
\label{floquet_effective}
\hat{H}_F^{(0)}
=
J_{\mathrm{eff}}
\sum_j
\big(
\hat{c}_j^\dagger \hat{c}_{j+1}
+
\text{H.c.}
\big)
+
\sum_j V_j \hat{c}_j^\dagger \hat{c}_j,
\end{equation}
with the renormalized hopping amplitude
\begin{equation}
\label{Jeff}
J_{\mathrm{eff}} = J\,J_0(\mathcal{K}),
\end{equation}
where $\mathcal{K} = K/\omega$ is the dimensionless ratio of drive strength to frequency.
Eq.~\eqref{floquet_effective} demonstrates that, in the high-frequency limit, the periodically driven system maps directly onto a static GPD model with a drive-renormalized hopping $J_{\mathrm{eff}}$. Consequently, at the zeros of the Bessel function $J_0(\mathcal{K}_n) = 0$ ($n = 1,2,\dots$), the effective tunneling vanishes ($J_{\mathrm{eff}} = 0$). In the absence of a quasiperiodic potential ($\lambda = 0$), this leads to standard dynamical localization accompanied by a complete collapse of the quasienergy band~\cite{Dynamical_localization,dyn_loc1,dyn_loc2,dyn_loc3,Photonic_ME}. At finite $\lambda$, the remaining onsite potential keeps all states strongly localized within the leading-order static framework, but also yields non-zero contributions to higher-order Magnus expansion terms~\cite{WELDPRR}. As the driving frequency is lowered, these higher-order terms become increasingly relevant, introducing new physical processes that shape both the spectral and dynamical properties of the system. Crucially, Eq.~\eqref{floquet_effective} forms the direct basis for mapping static mobility edge conditions to the driven regime.
\subsection{Analytical mobility edges}

The single particle eigenvalue equation corresponding to the effective Floquet Hamiltonian [Eq.~\eqref{floquet_effective}] can be expressed as
\begin{equation}
\label{eq:GAA_driven}
J_{\mathrm{eff}}\,[\psi_{j+1}+\psi_{j-1}]
+
\frac{2\lambda\cos(2\pi\alpha j+\phi)}
{1-\beta\cos(2\pi\alpha j+\phi)}\,
\psi_j
=
E\psi_j.
\end{equation}
Since the periodic drive enters solely through the renormalized hopping $J \to J_{\mathrm{eff}} = JJ_0(\mathcal{K})$, the analytical Floquet mobility edges follow directly from the static GPD results [Eqs.~\eqref{eq:static_ME_bounded} and \eqref{eq:static_ME_unbounded}] by replacing $J$ with $J_{\mathrm{eff}}$~\cite{GAAMOBILITYEDGE}. This yields the critical energy conditions
\begin{equation}
\label{Floquet_mobility_edge_cond}
\left|\frac{\beta E_c + 2\lambda}{JJ_0(\mathcal{K})}\right| =
\begin{cases}
2, & |\beta|<1, \\[6pt]
2|\beta|, & |\beta|\geqslant 1.
\end{cases}
\end{equation}
Solving for the critical energy $E_c$ yields the Floquet mobility edges
\begin{equation}
\label{Floquet_mobility_edges}
E_c =
\begin{cases}
\dfrac{\pm 2J\,J_0(\mathcal{K})-2\lambda}{\beta}, & |\beta|<1, \\[10pt]
\dfrac{\pm 2|\beta|\,J\,J_0(\mathcal{K})-2\lambda}{\beta}, & |\beta|\geqslant 1 .
\end{cases}
\end{equation}
A complete derivation of the Lyapunov exponent and energy spectrum using Avila’s global theory~\cite{Avila2015} is provided in Appendix~\ref{app:avila_full}. For $|\beta|<1$, the potential is bounded and $E_c$ delineates delocalized--localized (DL) mobility edges. Conversely, for $|\beta|\geqslant 1$, the unbounded potential eliminates fully extended states, causing $E_c$ to define multifractal--localized (ML) mobility edges.

Equation~\eqref{Floquet_mobility_edges} demonstrates that the spatial extent of the mobility edges is continuously tunable via the ratio $\mathcal{K} = K/\omega$. At the zeros of the Bessel function $J_0(\mathcal{K})=0$, the effective hopping vanishes ($J_{\mathrm{eff}}=0$), causing the mobility edges to collapse and yielding a completely localized spectrum. As $\mathcal{K}$ is varied between successive zeros of $J_0(\mathcal{K})$, the effective hopping is restored, leading to a repeated sequence of localization and mobility edge reemergence transitions. Periodic driving thus serves as a highly versatile knob to engineer mobility edges and tune spectral phase transitions purely through external driving parameters.
\section{Numerical verification of mobility edges and drive-induced localization}
\label{sec:numerics}
In this section, we numerically verify the analytically predicted mobility edges and the associated drive-induced localization phenomena discussed in Sec.~\ref{sec:floquet_ME}. Our analysis is based on exact diagonalization of the single particle Floquet operator $\hat{U}(T)=\mathcal{T}\exp\!\left[-i\int_0^T \hat{H}(t)\,dt\right]$, from which the quasienergies $\epsilon_n$ and Floquet eigenstates $|\psi_n\rangle$ are obtained via~\cite{Bukov04032015}
\[
\hat{U}(T) |\psi_n\rangle = e^{-i \epsilon_n T} |\psi_n\rangle,
\]
where $T$ is the time period of the drive.
We then perform a systematic study using complementary diagnostics, namely the fractal dimension $D_2$, the inverse participation ratio (IPR), and the spatial standard deviation of Floquet eigenstates. Together, these probes allow us to clearly distinguish localized, extended, and multifractal regimes in the driven system and to assess how periodic driving enables controlled engineering of mobility edges.
We consider $\omega = 10$ and $2$ as characteristic samples of the high-frequency and low-frequency regimes. In the high-frequency ($\omega=10$) regime, the effective Floquet Hamiltonian in Eq.~\eqref{High_frequency_approximated_H} is expected to be quantitatively reliable. In contrast, $\omega=2$ is used to explicitly demonstrate deviations from high-frequency behavior and highlight the role of finite-frequency corrections. Throughout this section, we employ periodic boundary conditions (PBC), primarily for convenience and to enable a direct comparison with the corresponding static model, where PBC is used. While, in principle, a driven system with a linear electric field is more naturally described using open boundary conditions (OBC), we have verified that the essential bulk physics remains unchanged under PBC. A comparison between PBC and OBC is provided in Appendix~\ref{app:boundary_conditions}. 
\begin{figure}
    \centering
    \includegraphics[width=0.48\textwidth]{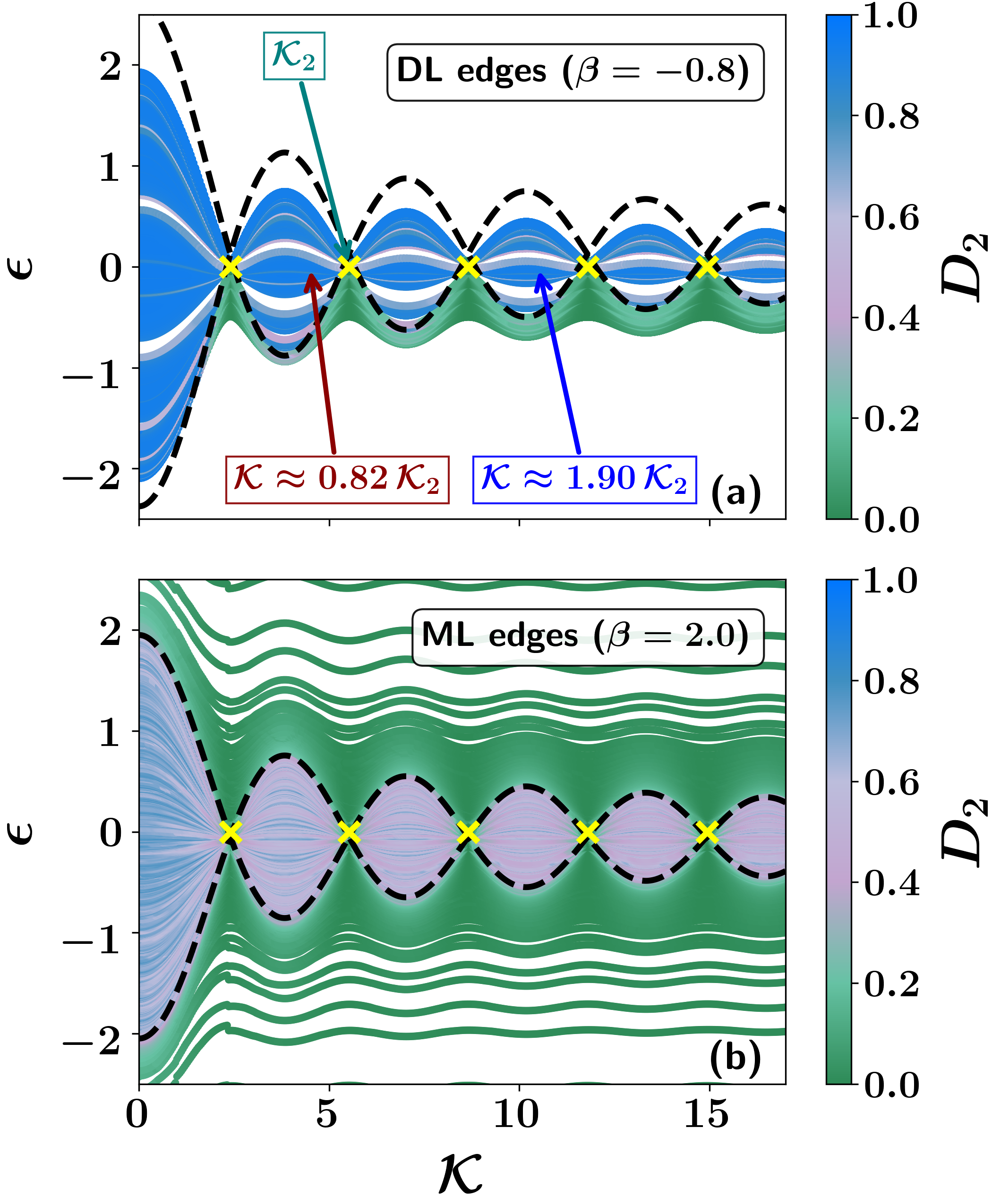}

\caption{
Fractal dimension $D_2$ of all the Floquet eigenstates of the driven GPD model at $\omega=10$, shown as a function of quasienergy $\epsilon$ and $\mathcal{K}$. 
(a) For $\beta=-0.8$, where periodic driving induces a DL edge in the Floquet spectrum. 
(b) For $\beta=2.0$, where the system exhibits an ML edge, with multifractal states ($0 < D_2 < 1$) in the central region and localized states near the spectral edges. 
Black dashed lines denote the analytically obtained mobility edges in Eq.~\eqref{Floquet_mobility_edges}. 
At the zeros of $J_0(\mathcal{K})$, occurring at $\mathcal{K} \approx 2.4048,\, 5.5201,\, 8.6537,\ldots$ (marked by yellow crosses), the effective hopping is suppressed, resulting in a fully localized spectrum. The arrows mark representative points whose spectral and dynamical properties are analyzed in detail in the following subsections. Here, $\mathcal{K}_2$ denotes the second zero of $J_0(\mathcal{K})$.
Results are shown for a system size $L = 987$, $\lambda=0.05$, and averaged over 10 realizations of the phase $\phi$.
}

    \label{fig:dynamic_ME}
\end{figure}
\subsection{Fractal dimension $D_2$ and Floquet mobility edge engineering}

The fractal dimension $D_q$ of a Floquet eigenstate is defined as~\cite{D2_Fit}
\begin{equation}
D_q = -\frac{1}{q-1}\lim_{L\to\infty}
\frac{\ln\!\left(\sum_{j=1}^{L} |\psi_{n,j}|^{2q}\right)}{\ln L},
\end{equation}
where $\psi_{n,j}$ denotes the amplitude of the $n^{\text{th}}$ normalized Floquet eigenstate at site $j$. In particular, for $q=2$, one obtains the standard fractal dimension previously defined in Eq.~\eqref{fractal_dimension}.
To identify Floquet mobility edges and characterize the nature of Floquet eigenstates across the quasienergy spectrum, we compute $D_2$.
The fractal dimension serves as a scale-invariant diagnostic of localization properties. For  extended states  $D_2=1$, for localized states $D_2=0$, and for multifractal states $0<D_2<1$.

\begin{figure*}[t]
\centering
\includegraphics[width=\textwidth]{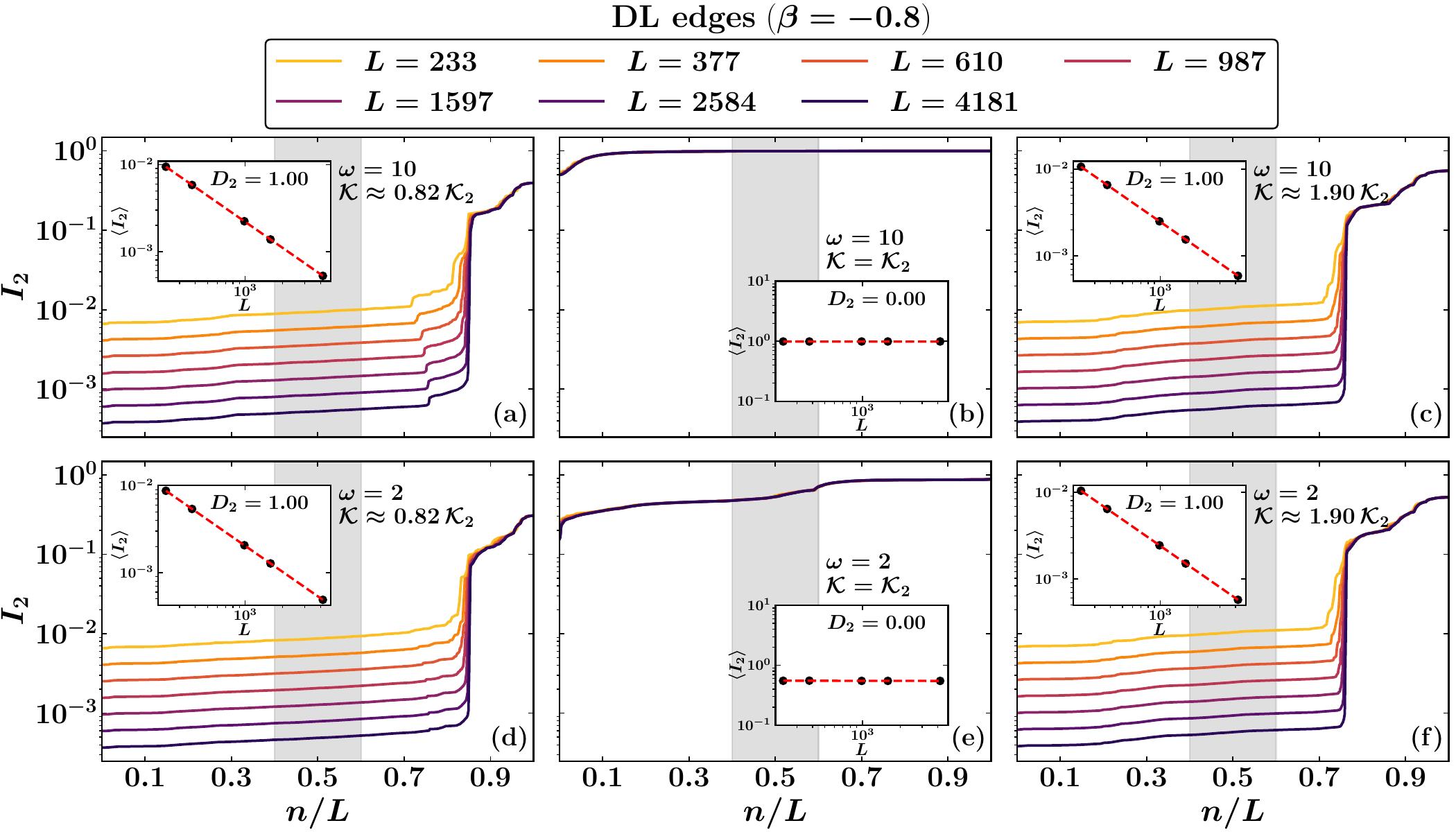}
\caption{
Sorted inverse participation ratio $I_2$ of Floquet eigenstates as a function of the normalized eigenstate index $n/L$ for different system sizes at $\beta=-0.8$ (with DL edges), shown for different values of $\mathcal{K}$, where $\mathcal{K}_2$ denotes the second zero of $J_0(\mathcal{K})$. (a)–(c) correspond to $\omega=10$, while (d)–(f) correspond to $\omega=2$. Insets show the average $\langle I_2 \rangle$ over Floquet eigenstates within the grey shaded window as a function of system size. The fractal exponent $D_2$ is extracted from linear fits of $\log \langle I_2 \rangle$ versus $\log L$, as defined in Eq.~\eqref{D_2_fit}. All results are for $\lambda=0.05$, averaged over 10 realizations of the phase $\phi$.
}
\label{fig:ipr_DDL}
\end{figure*}

The color maps of $D_2$ for $\omega=10$ as a function of quasienergy $\epsilon$ and driving parameter $\mathcal{K}$ in Fig.~\ref{fig:dynamic_ME} provide a global overview of the phase diagram of the Floquet spectrum for two representative values of $\beta$. For $\beta=-0.8$, corresponding to the bounded regime ($|\beta|<1$), the $D_2$ map clearly reveals DL edges [Fig.~\ref{fig:dynamic_ME}(a)]. Extended states with $D_2\simeq 1$ and localized states with $D_2\simeq 0$ coexist in the quasienergy spectrum, separated by sharp boundaries that evolve continuously as a function of $\mathcal{K}$. At the zeros of $J_0(\mathcal{K})$, the effective hopping vanishes ($J_{\mathrm{eff}} = 0$), and all Floquet eigenstates become localized across the spectrum. In contrast, for $\beta=2.0$, corresponding to the unbounded regime ($|\beta|\geqslant1$), the $D_2$ map displays a qualitatively different structure [Fig.~\ref{fig:dynamic_ME}(b)]. While a portion of the spectrum consists of localized states, the nonlocalized states exhibit intermediate values of $D_2$, indicating multifractal rather than  extended character, giving rise to ML edges. Again, at the zeros of $J_0(\mathcal{K})$, we observe localized states across the entire spectrum. 

In both the cases ($|\beta|<1$ and $|\beta|\geqslant1$), the analytical mobility edges obtained from Eq.~\eqref{Floquet_mobility_edges}, plotted as black dashed lines, are in excellent agreement with the exact numerics. 
However, this localization at the zeros of $J_0(\mathcal{K})$ is not accompanied by a complete collapse of the quasienergy band due to the presence of the onsite potential, and therefore does not correspond to true dynamical localization. The underlying reason is that higher-order terms in the Floquet--Magnus expansion continue to contribute, preventing complete suppression of transport, especially in the low-frequency regime where the high-frequency approximation breaks down.
\begin{figure*}[t]
\centering
\includegraphics[width=\textwidth]{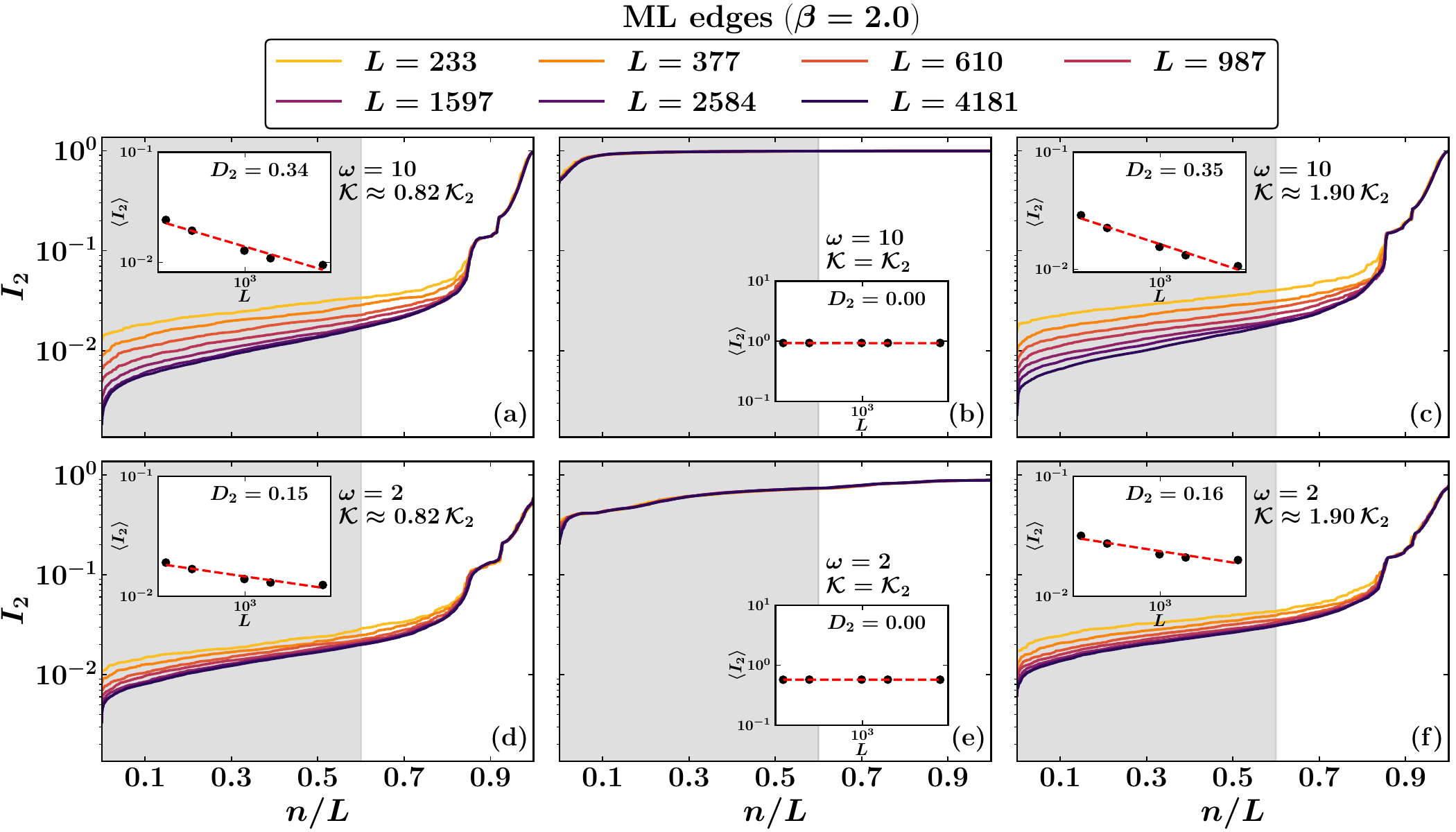}
\caption{
Sorted inverse participation ratio $I_2$ of Floquet eigenstates as a function of the normalized eigenstate index $n/L$ for different system sizes at $\beta=2.0$ (exhibiting ML edges), shown for different values of $\mathcal{K}$, where $\mathcal{K}_2$ denotes the second zero of $J_0(\mathcal{K})$. (a)-(c) correspond to $\omega=10$, while (d)-(f) correspond to $\omega=2$. Insets show the average $\langle I_2 \rangle$ over Floquet eigenstates within the grey shaded window as a function of system size. The fractal exponent $D_2$ is extracted from linear fits of $\log \langle I_2 \rangle$ versus $\log L$, as defined in Eq.~(\ref{D_2_fit}). All the results are for $\lambda=0.05$, and averaged over 10 realizations of the phase $\phi$.}
\label{fig:ipr_DML}
\end{figure*}
Moreover, the mobility edges form a sequence of bulging lobes that repeat as a function of $\mathcal{K}$. This structure arises from the periodic zeros and oscillatory behavior of the Bessel function. As $J_0(\mathcal{K})$ changes magnitude and sign, the effective bandwidth of the system expands and contracts, producing the observed periodic opening and closing of the mobility edge boundaries. By tuning the drive amplitude $K$, one can controllably shift and eliminate (at the zeros of $J_0(\mathcal{K})$) the mobility edges and generate different localization structures in the spectrum. 
At lower driving frequencies, particularly for $\omega \lesssim 1$, this well-defined mobility edge structure is progressively destroyed when $K$ is not sufficiently large. In this regime, strong quasienergy folding leads to significant overlap between Floquet bands, and higher-order terms in the Floquet--Magnus expansion become dominant. As a result, the dynamics is no longer governed primarily by the effective Hamiltonian in Eq.~\eqref{floquet_effective}, and the sharp mobility edge feature will give way to a more irregular structure.

Although the above results are obtained for a fixed $\lambda=0.05$, the evolution of the mobility edge structure and localization properties upon varying $\lambda$ is discussed in detail in Appendix~\ref{app:lambda_drive}.

\subsection{Inverse participation ratio and finite-size scaling}
To substantiate the above conclusions and quantitatively analyze the scaling behavior of Floquet eigenstates, we study the inverse participation ratio (IPR). The generalized IPR is defined as~\cite{PhysRevE.100.032117}
\begin{equation}
I_q(n)=\sum_{j=1}^{L} |\psi_{n,j}|^{2q},
\end{equation}
where $\psi_n$ denotes the $n^{\text{th}}$ Floquet eigenstate. For a system with Hilbert-space dimension $\mathcal{N}$, the generalized IPR scales as $I_q \sim \mathcal{N}^{-\tau_q}$, where $\tau_q = -\frac{\ln I_q}{\ln \mathcal{N}}$ is the scaling exponent. The fractal dimension $D_q$ is related to this exponent through $D_q = \tau_q/(q-1)$. In particular, for $q=2$ one obtains the usual inverse participation ratio
\begin{equation}
I_2(n)=\sum_{j=1}^{L} |\psi_{n,j}|^4.
\end{equation}
$I_2$ provides a direct finite-size probe of localization properties. It scales with system size as $I_2 \sim L^{-D_2}$, where the fractal dimension $D_2$ characterizes the nature of the eigenstates. $D_2 = 1$ for extended states, $D_2 = 0$ for localized states, and $0 < D_2 < 1$ for multifractal states.

Figs.~\ref{fig:ipr_DDL} and \ref{fig:ipr_DML} show the sorted $I_2$ as a function of the normalized eigenstate index $n/L$ for several system sizes, at representative values of $\mathcal{K} \approx 0.82\,\mathcal{K}_2,\;\mathcal{K}_2,$ and $\mathcal{K} \approx 1.90\,\mathcal{K}_2$, which are equivalent to $K/\omega \approx 4.5,\;5.52,$ and $10.5$, respectively. These points are indicated in Fig.~\ref{fig:dynamic_ME}(a). Here, $\mathcal{K}_2$ denotes the second zero of $J_0(\mathcal{K})$. Panels (a)-(c) in both figures correspond to $\omega=10$, while panels (d)-(f) correspond to $\omega=2$. The insets show $\langle I_2 \rangle$, averaged over eigenstates within the grey shaded region, plotted as a function of system size $L$ (for odd Fibonacci sizes) and used to extract the fractal dimension $D_2$ via~\cite{D2_Fit}
\begin{equation}
D_2 \approx -\frac{\log \langle I_2 \rangle}{\log L},
\label{D_2_fit}
\end{equation}
which follows from the finite-size scaling relation.
For $\beta=-0.8$, the $I_2$ data in Fig.~\ref{fig:ipr_DDL} exhibit a clear separation between localized and delocalized regimes. Away from $\mathcal{K}_2$, i.e., at $\mathcal{K} \approx 0.82\,\mathcal{K}_2$ and $\mathcal{K} \approx 1.90\,\mathcal{K}_2$ [panels (a), (c), (d), and (f)], the collapsed part of the spectrum shows size-independent $I_2$, indicating localized states, while the remaining part displays a systematic decrease of $I_2$ with increasing system size. Finite-size scaling of the averaged $I_2$ in the shaded region yields $D_2=1$ in the insets, confirming that these states are extended. At $\mathcal{K}_2$, the $I_2$ curves for different system sizes collapse almost perfectly across the entire spectrum [Fig.~\ref{fig:ipr_DDL}(b) and (e)], demonstrating drive-induced localization of all Floquet eigenstates.
 
For $\beta=2.0$ (Fig.~\ref{fig:ipr_DML}), the $I_2$ analysis reveals a qualitatively different scaling behavior. Away from $\mathcal{K}_2$, while localized states again form a size-independent plateau, the remaining part of the spectrum exhibits a slow power-law decay of $\langle I_2 \rangle$ with system size, signaling multifractal eigenstates. The extracted exponents yield finite values of $D_2$ between $0$ and $1$, in agreement with the multifractal character. The $I_2$ curves for different system sizes again collapse across the spectrum [Fig.~\ref{fig:ipr_DML}(b) and Fig.~\ref{fig:ipr_DML}(e)] at $\mathcal{K}_2$, showing the localized behaviour.

For $\mathcal{K} = \mathcal{K}_2$, in the high-frequency regime ($\omega = 10$), the values of $I_2$ are closer to $1$ for most eigenstates in both Fig.~\ref{fig:ipr_DDL} and Fig.~\ref{fig:ipr_DML} compared to $\omega = 2$, indicating stronger localization. Interestingly, away from $\mathcal{K}_2$, a comparison between the two driving frequencies shows that lowering the driving frequency enhances localization within the multifractal band. Although multifractal scaling persists, the decay of $I_2$ with system size becomes weaker at lower driving frequency. This is reflected in the smaller values of the extracted $D_2$ compared to the high-frequency case, as shown in the corresponding insets. This behavior is somewhat counterintuitive. In general, lowering the driving frequency enhances the role of higher-order Floquet corrections, which are often associated with increased heating and a tendency toward delocalization. In contrast, here we observe an opposite trend, with stronger localization emerging within the multifractal regime.
We attribute this behavior to higher-order Floquet corrections that become increasingly relevant at lower driving frequencies, as discussed in detail in Sec.~\ref{higher_oder_corrections}.

\begin{figure}[t]
\centering
\includegraphics[width=\linewidth]{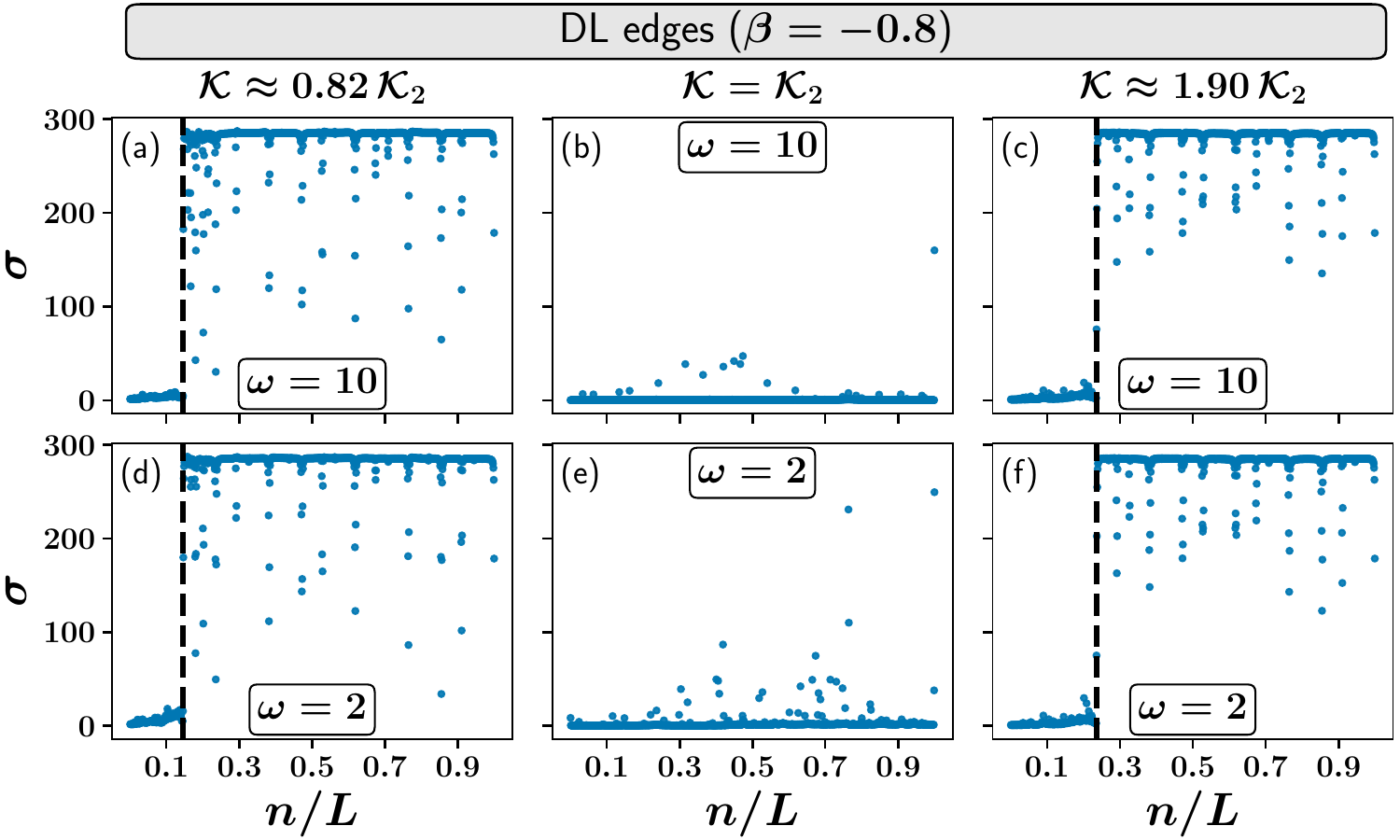}
\caption{
Standard deviation $\sigma$ of Floquet eigenstates as a function of the normalized eigenstate index $n/L$ for $\beta=-0.8$ (associated with DL edges), shown for different values of $\mathcal{K}$. 
(a)-(c) correspond to $\omega=10$, while (d)-(f) correspond to $\omega=2$. 
Black dashed lines denote the analytical mobility edge positions obtained from Eq.~\eqref{Floquet_mobility_edges}. 
At the Bessel zero $\mathcal{K} = \mathcal{K}_2$ [panels (b) and (e)], all states are localized. 
Away from this point [panels (a), (c), (d), and (f)], extended states dominate the spectrum, with localized states appearing near the lower-edge of the spectrum.  
Results are shown for a system size $L = 987$, $\lambda=0.05$, and averaged over 10 realizations of the phase $\phi$.
}
\label{fig:sigma_beta_neg}
\end{figure}

\subsection{Spatial extent of Floquet eigenstates}
\label{subsec:sigma_eigen}

To directly probe the spatial structure of Floquet eigenstates and to distinguish mobility edges in the quasienergy spectrum, we further analyze the standard deviation of each single particle Floquet eigenstate. For a normalized eigenstate $\psi_n$, the standard deviation is defined as~\cite{GAAMOBILITYEDGE}

\begin{equation}
\sigma(n) = \left[ \sum_{j=1}^{L} (j - \bar{j}_n)^2 \, |\psi_{n,j}|^2 \right]^{1/2},
\end{equation}
where the mean position for the $n^{\text{th}}$ eigenstate is given by
\begin{equation}
\bar{j}_n = \sum_j j \, |\psi_{n,j}|^2 .
\end{equation}
The quantity $\sigma$ serves as a direct and intuitive measure of spatial extent. Localized states are characterized by small values of $\sigma$ that remain independent of system size. Extended states exhibit large values of $\sigma$ that scale extensively with the system size. Multifractal states lie between these two limits and are distinguished by strong state-to-state fluctuations in $\sigma$, reflecting their nonergodic spatial structure.
\begin{figure}[t]
\centering
\includegraphics[width=\linewidth]{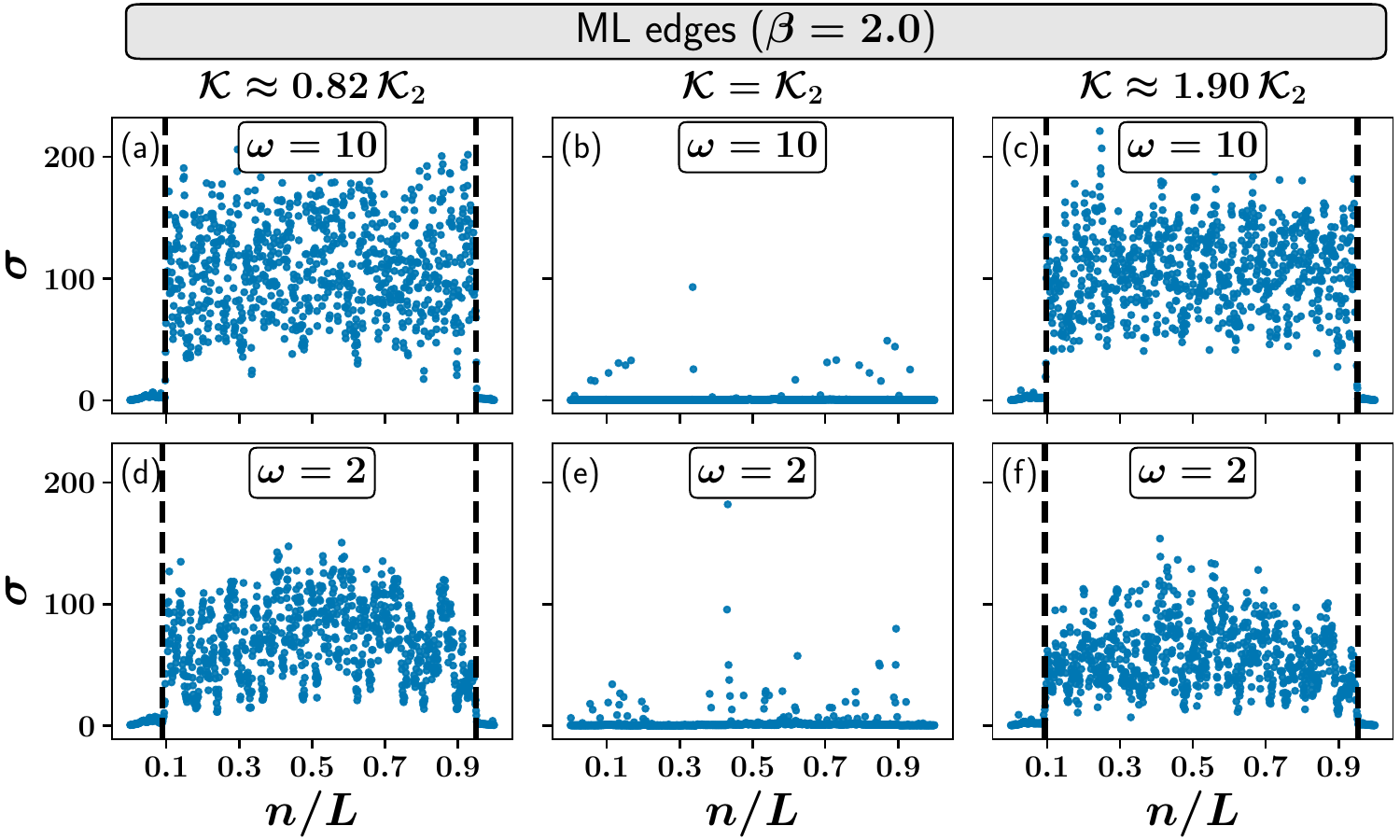}
\caption{
Standard deviation $\sigma$ of Floquet eigenstates as a function of the normalized eigenstate index $n/L$ for $\beta=2.0$ (associated with ML edges), shown for different values of $\mathcal{K}$. 
(a)-(c) correspond to $\omega=10$, while (d)-(f) correspond to $\omega=2$. 
Black dashed lines denote the analytical mobility edge positions obtained from Eq.~\eqref{Floquet_mobility_edges}. 
At the Bessel zero $\mathcal{K} = \mathcal{K}_2$ [panels (b) and (e)], all states are localized. 
Away from this point [panels (a), (c), (d), and (f)], multifractal dominate the spectrum, with localized states appearing near the lower-energy edge.  
Results are shown for a system size $L = 987$, $\lambda=0.05$, and averaged over 10 realizations of the phase $\phi$.
}
\label{fig:sigma_beta2}
\end{figure}

\begin{figure}[t]
    \centering
    \includegraphics[width=\linewidth]{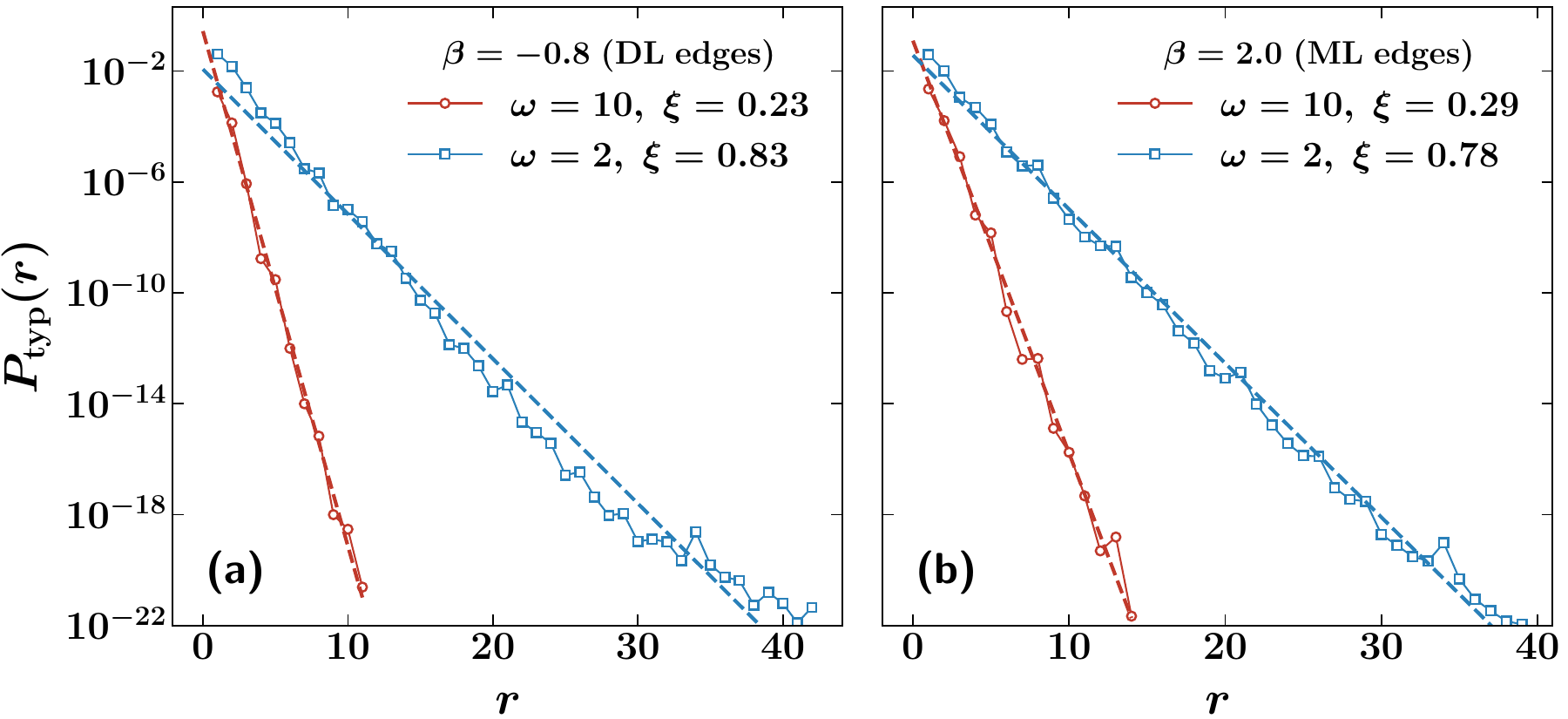}
    \caption{Typical wavefunction probability distribution $P_{\mathrm{typ}}(r)$ fitted against an exponential decay model $A e^{-r/\xi}$ at $\mathcal{K} = \mathcal{K}_2$. Panel (a) shows the decay profile for $\beta = -0.8$ (DL edges), while panel (b) shows the decay for $\beta = 2.0$ (ML edges) for driving frequencies $\omega = 10$ and $\omega = 2$. The localization length $\xi$ extracted from the fit is displayed inside the figure for the respective $\omega$ values.}
    \label{fig:typical_decay_K2}
\end{figure}

We analyze the standard deviation $\sigma$ of Floquet eigenstates for $\beta=-0.8$ and $\beta=2.0$, as shown in Figs.~\ref{fig:sigma_beta_neg} and \ref{fig:sigma_beta2}, respectively. Results for $\omega=10$ are presented in panels (a)-(c), while panels (d)-(f) correspond to $\omega=2$.
Away from the Bessel's zero $\mathcal{K}_2$, that is at $\mathcal{K}\approx 0.82\,\mathcal{K}_2$ and $\mathcal{K}\approx 1.90\,\mathcal{K}_2$ [panels (a), (c), (d), and (f)], the two regimes display distinct behavior. In the regime associated with DL edges ($\beta=-0.8$), most eigenstates exhibit large and nearly uniform $\sigma$, indicating extended states, with a small set of localized states with very small values of $\sigma$ appearing near the lower spectral edge. In contrast, for $\beta=2.0$, while localized states remain near the spectral edges, the central part of the spectrum shows moderate values of $\sigma$ with strong fluctuations, characteristic of multifractal eigenstates. Although these states are spatially extended, their support is highly inhomogeneous.
Consistent with the IPR results, at $\mathcal{K}=\mathcal{K}_2$ [panels (b) and (e) in Figs.~\ref{fig:sigma_beta_neg} and \ref{fig:sigma_beta2}], $\sigma \approx 0$ throughout the spectrum for both values of $\beta$, indicating drive-induced localization of all Floquet eigenstates. To quantitatively characterize the degree of localization across the spectrum, we compute the peak-aligned typical wavefunction profile defined as
\begin{equation}
P_{\mathrm{typ}}(r) = \exp \left[ \frac{1}{N} \sum_{n=1}^{N} \ln |\psi_n(j_n + r)|^2 \right],
\end{equation}
where $j_n$ denotes the peak position of the $n^{\text{th}}$ Floquet eigenstate, $r \ge 0$ is the relative spatial coordinate from the peak, and $N$ is the total number of states. As illustrated in Fig.~\ref{fig:typical_decay_K2} [panels (a) and (b) for $\beta = -0.8$ and $\beta = 2.0$, respectively], fitting the asymptotic exponential decay
\begin{equation}
P_{\mathrm{typ}}(r) = A \exp(-r/\xi)
\end{equation}
allows us to extract the spatial localization length $\xi$~\cite{RevModPhys.80.1355}, where larger (smaller) values of $\xi$ correspond to weaker (stronger) localization. For both values of $\beta$, the higher frequency ($\omega = 10$) yields a substantially smaller localization length $\xi$ compared to $\omega = 2$, directly demonstrating that higher driving frequencies induce stronger spatial confinement across the entire quasienergy spectrum.

The effect of increasing $\mathcal{K}$ away from $\mathcal{K}_2$ depends sensitively on the underlying regime. For $\beta=-0.8$, increasing $\mathcal{K}$ leads to a clear growth in the fraction of states with small $\sigma$, indicating enhanced localization. In contrast, for $\beta=2.0$, the fraction of multifractal states in the bulk remains largely unchanged. This difference can be traced back to the corresponding static model, where increasing $\lambda/J$ significantly enhances localization in the regime associated with DL edges [Fig.~\ref{fig:static_ME}(a)], while the regime associated with ML edges continues to support a substantial number of multifractal states even at larger $\lambda/J$ [Fig.~\ref{fig:static_ME}(b)]. The black dashed lines denote the analytically obtained mobility edge positions from Eq.~\eqref{Floquet_mobility_edges}, which are in good agreement with the numerical results. Further, for $\beta=2.0$, a comparison between the two driving frequencies shows that lowering $\omega$ enhances localization within the multifractal band, as reflected by smaller values of $\sigma$ at $\omega=2$ compared to $\omega=10$. This behavior can be attributed to higher-order Floquet corrections that become increasingly relevant at lower driving frequencies, consistent with the previous $I_2$ analysis. The next non-zero correction $\hat{H}_F^{(2)}$ is discussed in Sec.~\ref{higher_oder_corrections} to provide a qualitative understanding of these low-frequency effects.
\subsection{Effects of higher-order terms in the low-frequency regime}
\label{higher_oder_corrections}
To understand the localization properties of the Floquet eigenstates in the low-frequency regime (such as $\omega=2$), we examine the leading non-zero correction term in the effective Floquet Hamiltonian. For the parameters under consideration, lowering the frequency brings $\omega$ close to the single particle bandwidth, placing the system near a resonant regime where high-frequency asymptotic expansions may cease to be quantitatively exact. Nevertheless, evaluating the leading correction beyond zeroth order provides valuable qualitative insight into the emergence of physical processes that are entirely absent in the time-averaged Hamiltonian $\hat{H}_F^{(0)}$.
Using the Floquet--Magnus expansion, the effective Floquet Hamiltonian is systematically written as
\begin{equation}
\hat{H}_F = \hat{H}_F^{(0)} + \hat{H}_F^{(1)} + \hat{H}_F^{(2)} + \cdots.
\end{equation}
The zeroth-order contribution $\hat{H}_F^{(0)}$ is given by Eq.~\eqref{floquet_effective}. For the present model, the first-order correction vanishes, $\hat{H}_F^{(1)}=0$, making the second-order term $\hat{H}_F^{(2)}$ the leading correction beyond the time-averaged Hamiltonian:
\begin{equation}
\label{eq:second_order_correction}
\begin{aligned}
\hat{H}_F^{(2)} &=
\sum_{m\neq 0}
\frac{
J^2 J_m^2\!\left(\mathcal{K}\right)
}{
2(m\omega)^2
}
\Bigg[
\sum_j 2(V_{j+1}+V_{j-1}-2V_j)\hat{n}_j \\
&\quad
+ (-1)^m \sum_j (2V_{j+1}-V_j-V_{j+2})
(\hat{c}_j^\dagger \hat{c}_{j+2} + \text{H.c.})
\Bigg].
\end{aligned}
\end{equation}
The detailed derivation of $\hat{H}_F^{(2)}$ is provided in Appendix~\ref{app:low_freq}. Retaining terms up to second order, the effective Floquet Hamiltonian takes the compact form
\begin{equation}
\label{eq:HF_compact_final}
\begin{aligned}
\hat{H}_F \approx
\sum_j \Big[
& J_{\mathrm{eff}}
\big(
\hat{c}_j^\dagger \hat{c}_{j+1}
+ \text{H.c.}
\big)
+
J^{(2)}_j
\big(
\hat{c}_j^\dagger \hat{c}_{j+2}
+ \text{H.c.}
\big) \\
&+
\left( V_j + \tilde{V}_j \right)
\hat{c}_j^\dagger \hat{c}_j
\Big],
\end{aligned}
\end{equation}
where $J_{\mathrm{eff}} = J J_0(\mathcal{K})$ is the renormalized nearest-neighbor hopping amplitude. The second-order correction generates a position-dependent next-nearest-neighbor (NNN) hopping term with amplitude
\[
J^{(2)}_j =
\sum_{m\neq 0}
\frac{(-1)^m J^2 J_m^2(\mathcal{K})}{2(m\omega)^2}
\left(2V_{j+1}-V_j-V_{j+2}\right),
\]
and an additional onsite potential contribution
\[
\tilde{V}_j =
\sum_{m\neq 0}
\frac{J^2 J_m^2(\mathcal{K})}{(m\omega)^2}
\left(V_{j+1}+V_{j-1}-2V_j\right).
\]

The second-order correction introduces two competing physical
  mechanisms. First, the position-dependent NNN hopping processes
  $J^{(2)}_j$ tend to promote transport. This effect is most prominent
  near $\mathcal{K}=\mathcal{K}_2$, where the leading-order hopping
  $J_{\mathrm{eff}}$ vanishes, allowing $J^{(2)}_j$ to facilitate
  tunneling among otherwise localized states. Second, the correction
  renormalizes the onsite potential via $\tilde{V}_j$, which
  effectively corresponds to the discrete potential curvature
  $(V_{j+1}+V_{j-1}-2V_j)$. This potential modification acts as a
  local perturbation whose impact depends on the nature of the
  eigenstates. In particular, multifractal states characterized by
  strongly inhomogeneous spatial distributions are sensitive to such
  curvature and are consequently driven toward stronger
  localization. By contrast, delocalized states remain largely robust
  against these local potential variations. While other higher-order
  terms or multi-photon resonant processes may further quantitatively
  modify the spectrum at low frequencies, their explicit analytical
  calculation becomes cumbersome. Thus, $\hat{H}_F^{(2)}$ serves as
  the primary qualitative framework capturing the competing mechanisms
  that govern the observed low-frequency spectral and transport
  behaviors.

\section{Dynamics and transport across Floquet mobility edges}
\label{sec:Dynamics}
We now explore the dynamical properties of the Floquet Hamiltonian. In particular, we study transport along mobility edges and drive-induced localization points. To characterize the dynamics, we compute the root mean squared deviation (RMSD) and bipartite entanglement entropy. The dynamics is analyzed at stroboscopic times $t_n = nT$ ($T = 2\pi/\omega$), where an initial state $|\psi(0)\rangle$ evolves as
$|\psi(nT)\rangle = \left[ U(T) \right]^n |\psi(0)\rangle$ with $U(T) = \mathcal{T} \exp\left(-i \int_0^T H(t)\, dt \right)$. For this analysis, we consider a system of size $L = 987$ and $\lambda=0.05$ with open boundary conditions.
\subsection{Root-mean-squared deviation (RMSD)}
\label{subsec:RMSD}

To characterize the dynamics and the nature of transport in the periodically driven system, we analyze the spreading of a single particle wave packet initially localized at the center of the one-dimensional chain. The initial state is chosen as
\begin{equation}
\ket{\psi(0)} = \ket{j_0}, \qquad j_0 = L/2 ,
\label{initial_state}
\end{equation}
where all sites are empty except $j_0$.
The root-mean-squared deviation is defined as~\cite{Vatsana_dyn}
\begin{equation}
X(t) = \left[ \sum_{j=0}^{L} (j - j_0)^2 |\psi_j(t)|^2 \right]^{1/2},
\label{eq:RMSD}
\end{equation}
where $|\psi_j(t)|^2$ denotes the probability density at site $j$ at time $t$.
At long times (after disregarding initial transients and before saturation), the RMSD typically follows a power-law scaling
\begin{equation}
X(t) \sim t^{\gamma},
\label{eq:RMSD_GROWTH}
\end{equation}
where the exponent $\gamma$ characterizes the nature of transport. Ballistic transport corresponds to $\gamma \simeq 1$, diffusive transport denotes $\gamma \simeq 1/2$, while $0 < \gamma < 1/2$ indicates subdiffusive spreading, and $1/2 < \gamma < 1$ indicates superdiffusive spreading~\cite{Supplementary}. In the localized regime, $X(t)$ saturates at very small values, corresponding to $\gamma \simeq 0$.
In systems with single particle mobility edges, the dynamics become intrinsically heterogeneous. Since localized, extended, and multifractal eigenstates coexist in the spectrum, the spreading of an initially localized wave packet reflects a competition between these components, leading to intermediate dynamical regimes with superdiffusive transport for DL edges and subdiffusive transport for ML edges~\cite{Supplementary,D2_Fit, PhysRevLett.69.695, PhysRevLett.79.1959, JPSJ.66.314}.

\begin{figure}[t]
\centering
\includegraphics[width=\columnwidth]{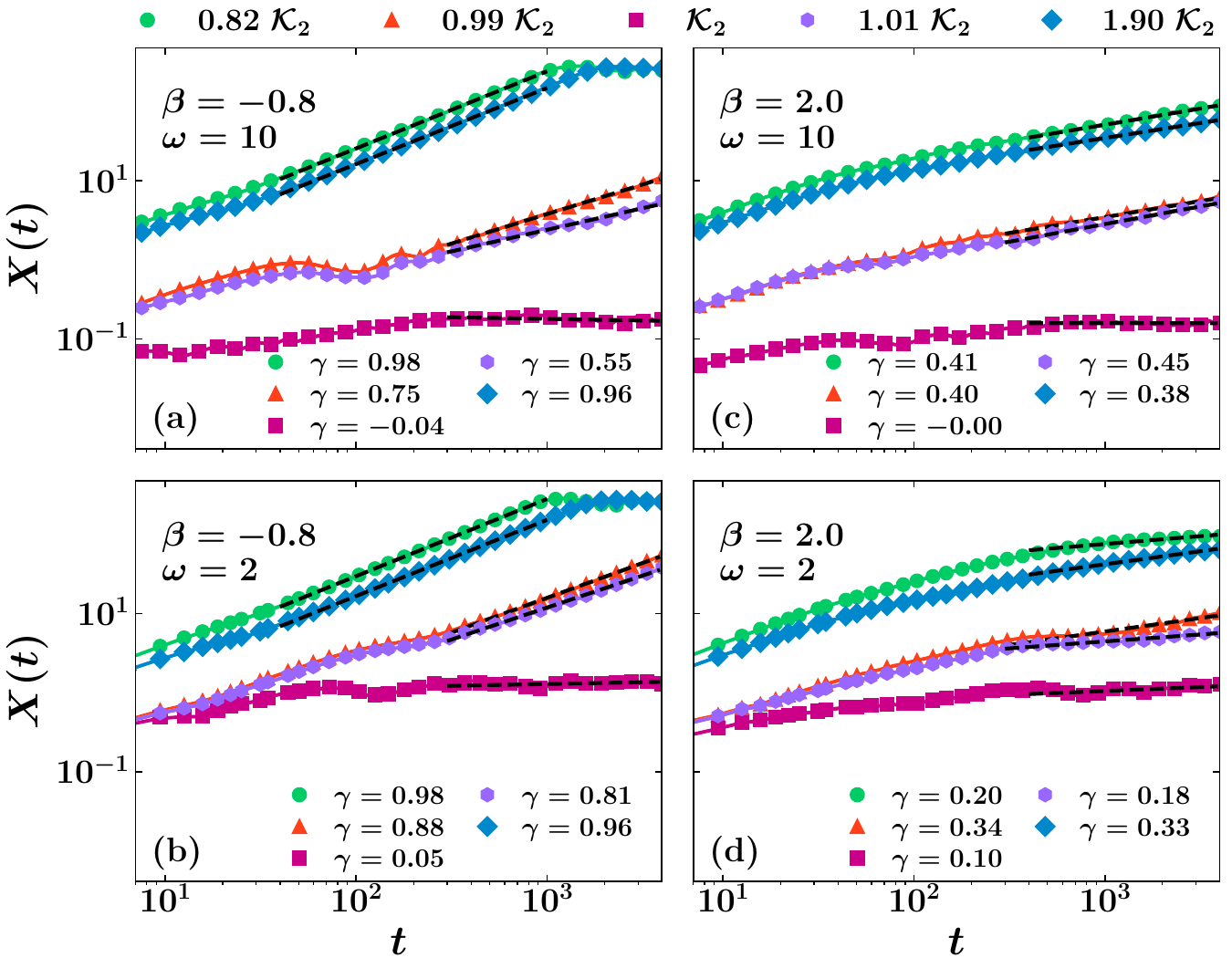}
\caption{
Dynamics of the root-mean-squared displacement $X(t)$ of a wave packet initially localized at the center of the chain. Rows correspond to different values of the driving frequency $\omega$, while columns correspond to different values of $\beta$. Different values of $\mathcal{K}$ are represented by distinct markers and colors, with $\mathcal{K}_2 \approx 5.52$ denoting the second zero of the Bessel function $J_0(\mathcal{K})$. Black dashed lines indicate power-law fits of $X(t) \sim t^{\gamma}$ in the long-time regime, after initial transients and before saturation. The extracted growth exponents $\gamma$, corresponding to each value of $\mathcal{K}$, are indicated by the respective markers. Results are shown for a system size $L = 987$, $\lambda=0.05$, and averaged over 100 realizations of the phase $\phi$.
}
\label{fig:RMSD}
\end{figure}

We study the dynamics of $X(t)$ for $\beta=-0.8$, associated with the presence of DL edges, for driving frequencies $\omega=10$ and $2$, as shown in Fig.~\ref{fig:RMSD}(a) and Fig.~\ref{fig:RMSD}(b). The transport exponent $\gamma$ is obtained by fitting the late-time behavior of $X(t)$ to the scaling relation in Eq.~\eqref{eq:RMSD_GROWTH}, as indicated by the black dashed lines in the figures.
Away from the drive-induced localization point $\mathcal{K}_2$, i.e., at $\mathcal{K}\approx 0.82\,\mathcal{K}_2$ and $\mathcal{K}\approx 1.90\,\mathcal{K}_2$, the Floquet spectrum is dominated by delocalized states, as established from the spectral analysis. The dynamics is therefore governed by these states, resulting in superdiffusive to nearly ballistic spreading of the wave packet for both driving frequencies. The extracted exponent $\gamma$ remains nearly unchanged between $\omega=10$ and $\omega=2$.
In the vicinity of $\mathcal{K}_2$, for $\mathcal{K}\approx 0.99\,\mathcal{K}_2$ and $\mathcal{K}\approx 1.01\,\mathcal{K}_2$, the increasing fraction of localized states strongly influences the transport. As a result, the growth exponent $\gamma$ is reduced, and the spreading becomes more clearly superdiffusive, with an extended power-law regime in time. Lowering the driving frequency further enhances the growth exponent in this regime (near $\mathcal{K}_2$), reflecting the influence of higher-order Floquet processes in the presence of a large fraction of localized states, although the transport remains superdiffusive.
At $\mathcal{K}_2$, the spreading of the wave packet is strongly suppressed, with $\gamma \approx 0$ for both driving frequencies, indicating drive-induced localization. However, the saturation value of $X(t)$ is higher for $\omega=2$ than for $\omega=10$, reflecting weaker localization at lower frequency. This difference arises from enhanced transient spreading at early times due to the higher-order processes.

\begin{figure}[t]
\centering
\includegraphics[width=\columnwidth]{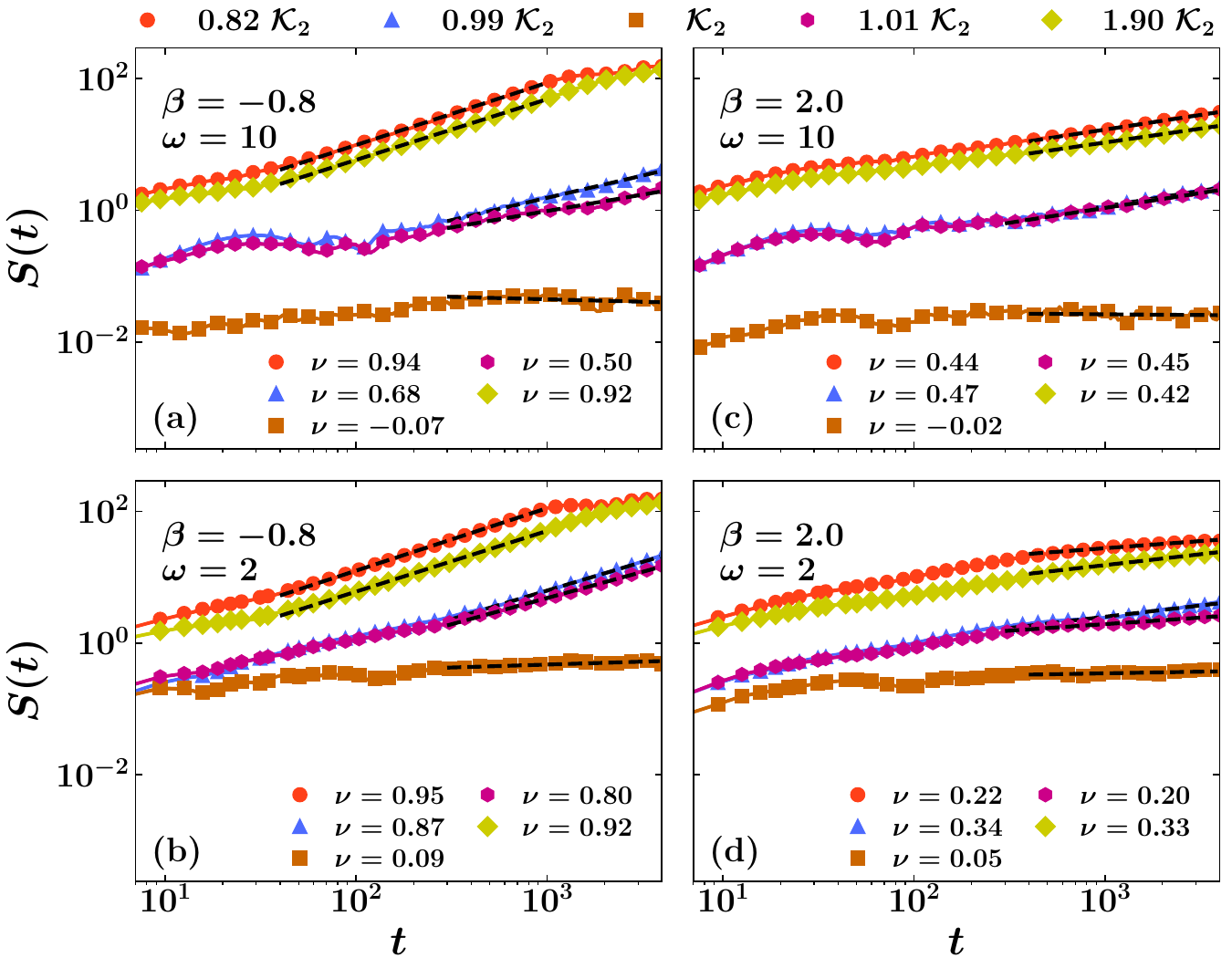}
\caption{
Dynamics of the half-chain entanglement entropy $S(t)$ for the periodically driven GPD model. Rows correspond to different values of the driving frequency $\omega$, while columns correspond to different values of $\beta$. Different values of $\mathcal{K}$ are represented by distinct markers and colors, with $\mathcal{K}_2 \approx 5.52$ denoting the second zero of the Bessel function $J_0(\mathcal{K})$. Black dashed lines denote power-law fits of $S(t) \sim t^{\nu}$ in the long-time regime, after initial transients and before saturation. The extracted growth exponents $\nu$, corresponding to each value of $\mathcal{K}$, are indicated by the respective markers. Results are shown for a system size $L = 987$, $\lambda=0.05$, and averaged over 100 realizations of the phase $\phi$.
}
\label{fig:EE}
\end{figure}
We now consider $\beta=2.0$, corresponding to the regime which hosts ML edges, shown in Fig.~\ref{fig:RMSD}(c) and Fig.~\ref{fig:RMSD}(d). In this case, the Floquet spectrum consists of multifractal and localized eigenstates. Since multifractal states are not fully ergodic and exhibit inhomogeneous spatial structure, they give rise to anomalous transport characterized by subdiffusive spreading of the wave packet.
Away from $\mathcal{K}_2$, the transport remains subdiffusive for both driving frequencies, reflecting the dominance of multifractal states in the bulk of the spectrum. In the vicinity of $\mathcal{K}_2$, the increased fraction of localized states further suppresses transport, leading to smaller values of $X(t)$ over the accessible time window. Lowering the driving frequency leads to a clear reduction in the growth exponent $\gamma$, indicating enhanced localization tendencies. This behavior is consistent with the spectral analysis and can be attributed to higher-order Floquet processes that induce effective potential modifications, suppressing transport within the multifractal band.
At $\mathcal{K}=\mathcal{K}_2$, the system again exhibits strong suppression of spreading, which is more resolved at the higher frequency value. Overall, while the full numerical dynamics incorporates all-order contributions, the analytical correction $\hat{H}_F^{(2)}$ captures key physical processes such as emergent next-nearest-neighbor hoppings and potential modifications that provide qualitative insight into these low-frequency dynamical behaviors (Sec.~\ref{higher_oder_corrections}).
\subsection{Entanglement entropy}
\label{subsec:EE}

In addition to real-space wave-packet spreading, we further characterize the dynamical regimes through the growth of the half-chain entanglement entropy. While the RMSD measures the spatial spreading of particles, the entanglement entropy captures the build-up of quantum correlations between spatial subsystems during time evolution. As such, it provides a complementary probe of transport processes and supports our findings from the RMSD analysis.
The system is initialized in a N\'eel state,
\begin{equation}
\ket{\psi(0)} = \prod_{j=1}^{L/2} \hat{c}_{2j-1}^{\dagger}\ket{0},
\end{equation}
where fermions occupy every alternate lattice site. This highly ordered product state contains no initial entanglement between the two halves of the system. Any subsequent growth of entanglement entropy therefore arises from correlations generated by the periodic driving.

The entanglement entropy of subsystem $A$ (taken to be the left half of the chain) is defined as~\cite{NielsenChuang2010}
\begin{equation}
S(t) = -\mathrm{Tr}\!\left[\hat{\rho}_A(t)\ln\hat{\rho}_A(t)\right],
\end{equation}
where the reduced density matrix of subsystem $A$ is obtained by tracing out the degrees of freedom of the complementary subsystem $B$, i.e. $\hat{\rho}_A(t)=\mathrm{Tr}_B\,\hat{\rho}(t)$. For the quadratic fermionic system studied here, the entanglement entropy can be efficiently computed from the eigenvalues $\{\Lambda_k(t)\}$ of the single particle correlation matrix of subsystem $A$~\cite{Peschel_2003,Peschel_2009}. The correlation matrix is defined as
$\hat{C}_{mn}^{A}(t)=\langle \psi(t)|\hat{c}_m^\dagger \hat{c}_n|\psi(t)\rangle$,
where $m,n \in A$, and the state evolves as $|\psi(t)\rangle = e^{-i\hat{H}t}|\psi(0)\rangle$. Diagonalizing $\hat{C}^{A}$ yields the eigenvalues $\{\Lambda_k(t)\}$, from which the von Neumann entropy is given by
\begin{equation}
S(t) = -\sum_k \left[(1-\Lambda_k(t))\ln(1-\Lambda_k(t))+\Lambda_k(t)\ln \Lambda_k(t)\right].
\end{equation}
This formulation allows the many-body entanglement entropy to be computed directly from the correlation matrix of the underlying non-interacting Hamiltonian. We study the time evolution of the half-chain entanglement entropy for the same parameter sets used in the RMSD analysis and fit the entanglement entropy as
\begin{equation}
\label{entropy_fit}
S(t) \sim t^{\nu},
\end{equation}
where $\nu$ is the growth exponent. The growth of $S(t)$ directly reflects the underlying transport mechanisms discussed in the previous subsection. In particular, it is closely related to particle-number fluctuations across the bipartition~\cite{PhysRevB.97.125116}.

$S(t)$ is analyzed as a function of time in Fig.~\ref{fig:EE}, using the same values of $\mathcal{K}$ as in the RMSD analysis. The fitted curves, indicated by black dashed lines, are used to extract the growth exponent $\nu$ using the scaling relation in Eq.~\eqref{entropy_fit}.
We first consider the regime associated with DL edges ($\beta=-0.8$), shown in Fig.~\ref{fig:EE}(a) and Fig.~\ref{fig:EE}(b) for $\omega=10$ and $2$, respectively. Away from the drive-induced localization point $\mathcal{K}_2$, $S(t)$ exhibits a clear power-law growth. In this regime, the entanglement entropy shows rapid growth, ranging from superdiffusive to nearly ballistic behavior, consistent with the dominance of delocalized states.
In the vicinity of $\mathcal{K}_2$, i.e., at $\mathcal{K} = 0.99\,\mathcal{K}_2$ and $1.01\,\mathcal{K}_2$, the growth exponent $\nu$ is reduced, reflecting the increasing influence of localized states. The resulting dynamics remains superdiffusive but becomes slower in time. Lowering the driving frequency ($\omega=2$) enhances the growth exponent in this regime, reflecting higher-order Floquet processes that lead to a stronger buildup of correlations.
We now turn to the regime associated with ML edges ($\beta=2.0$), shown in Fig.~\ref{fig:EE}(c) and Fig.~\ref{fig:EE}(d). Away from $\mathcal{K}_2$, the entanglement entropy exhibits subdiffusive growth across the considered values of $\mathcal{K}$, reflecting the presence of multifractal eigenstates. The values of $S(t)$ are comparatively lower for $\mathcal{K}$ in the vicinity of $\mathcal{K}_2$ due to the increased fraction of localized states. Lowering the driving frequency ($\omega=2$) further reduces the growth exponent, indicating enhanced localization tendencies in the multifractal states driven by higher-order contributions, in agreement with the spectral and RMSD analyses. At $\mathcal{K}=\mathcal{K}_2$, the growth of $S(t)$ is strongly suppressed for both values of $\beta$, with the entanglement entropy approaching saturation at long times, indicative of drive-induced localization. However, at the lower frequency ($\omega=2$), $S(t)$ attains higher values and exhibits a small but finite growth, signaling weaker localization. This behavior is consistent with the RMSD analysis.

The close qualitative agreement between the entanglement entropy and the RMSD confirms that the transport properties and entanglement growth are governed by the same underlying structure of the Floquet eigenstates. To further substantiate this picture, we have also analyzed the return probability of an initially localized wavepacket as a function of time in Appendix~\ref{app:return_prob}. Additionally, a detailed quantitative analysis confirming the stability of the extracted exponents $\gamma$ and $\nu$ against variations in fitting windows and system sizes is presented in Appendix~\ref{app:exponent_stability}.
\section{Conclusion}
\label{sec:conclusion}
In this work, we investigate the periodically driven Ganeshan-Pixley-Das Sarma (GPD) model subjected to a linear electric field drive and demonstrate the emergence of mobility edges in its Floquet spectrum. These mobility edges closely resemble those of the static model, indicating that the underlying localization properties persist under periodic driving, while their structure can be tuned via the driving frequency and amplitude. In addition, as expected from coherent destruction of tunneling at the zeros of the Bessel function, the Floquet spectrum becomes strongly localized in the high-frequency regime, leading to significantly suppressed transport and enhanced memory of the initial state. Applying Avila's global theory to the effective Hamiltonian yields analytical expressions for these Floquet mobility edges, which show excellent agreement with the numerical results.

To probe the dynamical consequences, we analyze the root-mean-squared
deviation (RMSD), entanglement entropy, and return probability. Across
the DL mobility edge in the $|\beta|<1$ regime, the dynamics shows
superdiffusive to nearly ballistic spreading, which is progressively
reduced near the drive-induced localization points. In contrast, near
the ML edge for $|\beta| \geqslant 1$, the dynamics is governed by
multifractal and localized eigenstates and exhibits subdiffusive
behavior with smaller growth exponents. These dynamical signatures are
consistent with the spectral analysis based on IPR, $D_2$, and the
standard deviation of Floquet eigenstates, and the spatial localization length $\xi$. Interestingly, within a
single lobe bounded by two successive zeros of the Bessel function,
the fraction of localized states evolves from close to unity to
progressively smaller values and then increases again, implying that
tuning the drive amplitude $K$ alone enables controlled access to
different transport regimes. Consequently, the extent of
superdiffusive and subdiffusive transport in the respective parameter
regimes can be continuously tuned using $K$ as a robust control
knob. We further find that lowering the driving frequency enhances
higher-order Floquet processes, leading to deviations from the
high-frequency regime and revealing behavior that deviates from
conventional expectations. A particularly counter intuitive phenomenon
we uncover is that of enhanced localization within the multifractal
band in the low-frequency regime.

Taken together, our results establish the periodically driven GPD model as a controllable platform for realizing Floquet mobility edges and provide a comprehensive picture of transport across different regimes, highlighting the potential of Floquet engineering for manipulating localization in quasiperiodic systems. Our work also provides a route toward experimental realization of such models in the noninteracting limit~\cite{WELDPRR,WELDPRX,sunil2026localizationhoppingdisorderquasiperiodic}, and opens avenues to explore the fate of Floquet-engineered mobility edges in the presence of interactions and their role in thermalization dynamics.

\section*{Acknowledgments}
We acknowledge the High Performance Computing (HPC) facility at IISER
Bhopal, where the large-scale calculations for this project were
performed. We thank David Weld and Sayan Choudhury and Arnab Das for
stimulating discussions and insightful comments. M.K.\ acknowledges
support from the Council of Scientific and Industrial Research (CSIR),
India, through a PhD fellowship.  \appendix
\section{Transfer matrix decomposition and Lyapunov exponent}
\label{app:avila_full}

In this Appendix, we demonstrate the detailed derivation of mobility edges using Avila's theory.  We start by representing our model [Eq.~\eqref{eq:GAA_driven}] in the transfer matrix form as
\begin{equation*}
\Psi_{j+1}=T_j \Psi_j
\end{equation*}
where $\Psi_j^T=(\psi_{j} \hspace{2mm} \psi_{j-1})^T$. $T_j$ is the transfer matrix given by
\begin{equation}
T_j =
\begin{pmatrix}
\frac{E}{J_{\mathrm{eff}}}-\frac{2\lambda}{J_{\mathrm{eff}}}\frac{\cos(2\pi\alpha j+\phi)}
{1-\beta\cos(2\pi\alpha j+\phi)} & -1 \\
1 & 0
\end{pmatrix}.
\label{transfer_matrix}
\end{equation}
Using the transfer matrix, one can compute the Lyapunov exponent defined as,
\begin{equation}
\gamma(E) = \lim_{L \to \infty} \frac{1}{2\pi L} \int \ln \| T_L(\phi) \| \, d\phi,
\end{equation}
where
\begin{equation*}
T_L = \prod_{j=1}^{L} T_j,
\end{equation*}
and $\|T_L\|$ denotes the norm of the matrix $T_L$. To use Avila's theory~\cite{Avila2015}, we remove the singularity in the transfer matrix by factorizing it as
\begin{equation}
T_j = A_j B_j,
\end{equation}
with
\begin{align*}
 A_j = \frac{1}{1-\beta \cos(2\pi \alpha j + \phi)}, \quad
 B_j =
\begin{pmatrix}
B_{11} & B_{12} \\
B_{21} & 0
\end{pmatrix},
  \label{T_decompose}
\end{align*}
where
$B_{11}=
\frac{E}{J_{\mathrm{eff}}}\left[1-\beta \cos(2\pi \alpha j + \phi)\right]
-\frac{2\lambda}{J_{\mathrm{eff}}}\cos(2\pi \alpha j + \phi)$ and 
$B_{21} = -B_{12} = 1-\beta \cos(2\pi \alpha j + \phi)$.
This decomposition separates the Lyapunov exponent into two contributions,
\begin{equation}
\gamma(E) = \gamma_A(E) + \gamma_B(E)
\label{Total_gamma_def}
\end{equation}
where
\begin{equation*}
\gamma_A(E)
=
\lim_{L\to\infty}
\frac{1}{2\pi L}
\int
\ln \left\lVert A_L(\phi) \right\rVert
\, d\phi,
\end{equation*}
with $A_L(\phi) = \prod_{j=1}^{L} A_j(\phi)$. Then by using classical Jensen's formula and ergodic theory, we can write~\cite{PhysRevB.100.125157, gradshteyn_ryzhik_2007},
\begin{align}
\gamma_A(E)
&=
\frac{1}{2\pi}
\int_0^{2\pi}
\ln
\left(
\frac{1}{|1-\beta \cos\phi|}
\right)
d\phi \nonumber \\
          &=
        \begin{cases}
-\ln \left( \dfrac{1+\sqrt{1-\beta^2}}{2} \right), & |\beta| < 1, \\[8pt]
-\ln \left( \dfrac{|\beta|}{2} \right), & |\beta| \ge 1.
\end{cases}.
        \label{gamma_A}
\end{align}
Further, we use Avila's theory to calculate $\gamma_B$ as~\cite{PhysRevB.103.174205}
\begin{equation}
\label{gamma_B_def}
\gamma_B(E)
=
\lim_{L\to\infty}
\frac{1}{2\pi L}
\int_0^{2\pi}
\ln \left\lVert B_L(\phi) \right\rVert
\, d\phi
\end{equation}
with $B_L(\phi) = \prod_{j=1}^{L} B_j(\phi)$.
A key step is to complexify the phase, $\phi \rightarrow \phi + i\eta$. In the asymptotic limit $\eta \to \infty$, the transfer matrix simplifies to
\begin{equation*}
B_j(\phi) 
= e^{-i(2\pi \beta j + \phi)} \, \frac{e^{\eta}}{2}
\begin{pmatrix}
-\dfrac{\beta E + 2\lambda}{J_{\mathrm{eff}}} & -\beta \\
\beta & 0
\end{pmatrix}.
\end{equation*}
Defining $P = \frac{\beta E + 2\lambda}{J_{\mathrm{eff}}}$, the norm of $B_j$ can be written as
\begin{equation*}
\left\lVert B_j \right\rVert \sim e^{\eta} \, \frac{\left| P + \sqrt{P^2 - 4\beta} \right|}{4}.
\end{equation*} 
Substituting into Eq.~\eqref{gamma_B_def}, we obtain
\begin{equation}
\gamma_B(E,\eta)=
\begin{cases}
\displaystyle
\eta + \ln\!\left(\frac{|P|+\sqrt{P^2-4\beta^2}}{4}\right),
& P^2>4\beta^2, \\[8pt]
\displaystyle
\eta + \ln\!\left(\frac{|\beta|}{2}\right),
& P^2<4\beta^2.
\end{cases}
\end{equation}
The slope of $\gamma_B(E,\eta)$ is $1$ in the limit $\eta \to \infty$. However, since $\det B_j \neq 1$, the cocycle is not in $SL(2,\mathbb{R})$. In this case, following Refs.~\cite{Jitomirskaya2013, Jitomirskaya2017, kumar2026}, $\gamma_B$ is a convex, piecewise-linear function of $\eta$ with quantized slopes taking integer and half-integer values.  
As $\eta$ is decreased from infinity, the slope near $\eta=0$ can therefore take values $1$, $1/2$, or $0$. According to Avila’s global theory, an energy $E$ lies outside the spectrum if and only if $\gamma_B(E,0)>0$ and $\gamma_B(E,\eta)$ remains flat in a neighborhood of $\eta=0$. For the present GPD model, numerical evidence indicates the absence of states corresponding to slope $1/2$\footnote{If states with slope $1/2$ were present, the above analytical expressions would no longer be valid, and one would expect additional localized states to appear between the mobility edges~\cite{kumar2026}. Such features are not observed in our numerical results [Fig.~\ref{fig:dynamic_ME}].}. Consequently, in the vicinity of $\eta=0$, $\gamma_B$ reduces to
\begin{equation}
\label{gamma_B}
\gamma_B(E)=
\begin{cases}
\displaystyle
\ln\!\left(\frac{|P|+\sqrt{P^2-4\beta^2}}{4}\right),
& P^2>4\beta^2, \\[8pt]
\displaystyle
 \ln\!\left(\frac{|\beta|}{2}\right),
& P^2<4\beta^2.
\end{cases}
\end{equation}
Hence, total LE can be written as
\begin{equation}
\gamma(E)
=
\max\{0,\,\gamma_A(E)+\gamma_B(E,0)\}.
\end{equation}
Using Eq.~\eqref{gamma_A} and ~\eqref{gamma_B}  we get 
$\gamma(E)
=
\max\{0,\,F_\beta(P)\}$,
where 
\begin{equation}
F_\beta(P)=
\begin{cases}
\displaystyle
\ln\!\Bigg(
\frac{|P|+\sqrt{P^2-4\beta^2}}
{2\big(1+\sqrt{1-\beta^2}\big)}
\Bigg),
& |\beta|<1,\; P^2>4\beta^2, \\[8pt]
0,
& |\beta|<1,\; P^2<4\beta^2, \\[8pt]
\displaystyle
\ln\!\Bigg(
\frac{|P|+\sqrt{P^2-4\beta^2}}
{2|\beta|}
\Bigg),
& |\beta|\geqslant1,\; P^2>4\beta^2, \\[8pt]
0,
& |\beta|\geqslant1,\; P^2<4\beta^2 .
\end{cases}
\end{equation}
The mobility edges, i.e., the critical energies corresponding to the vanishing of the Lyapunov exponent, are given by
\begin{equation}
E_c =
\begin{cases}
\dfrac{2J\,J_0(\mathcal{K})-2\lambda}{\beta}, & |\beta|<1, \\[8pt]
\dfrac{2|\beta|\,J\,J_0(\mathcal{K})-2\lambda}{\beta}, & |\beta|\ge 1 
\end{cases}
\end{equation}
which corresponds to Eq.~\eqref{Floquet_mobility_edges} in the main text.

\section{Mobility edges under different boundary conditions}
\label{app:boundary_conditions}
\begin{figure}
    \centering
    \includegraphics[width=0.48\textwidth]{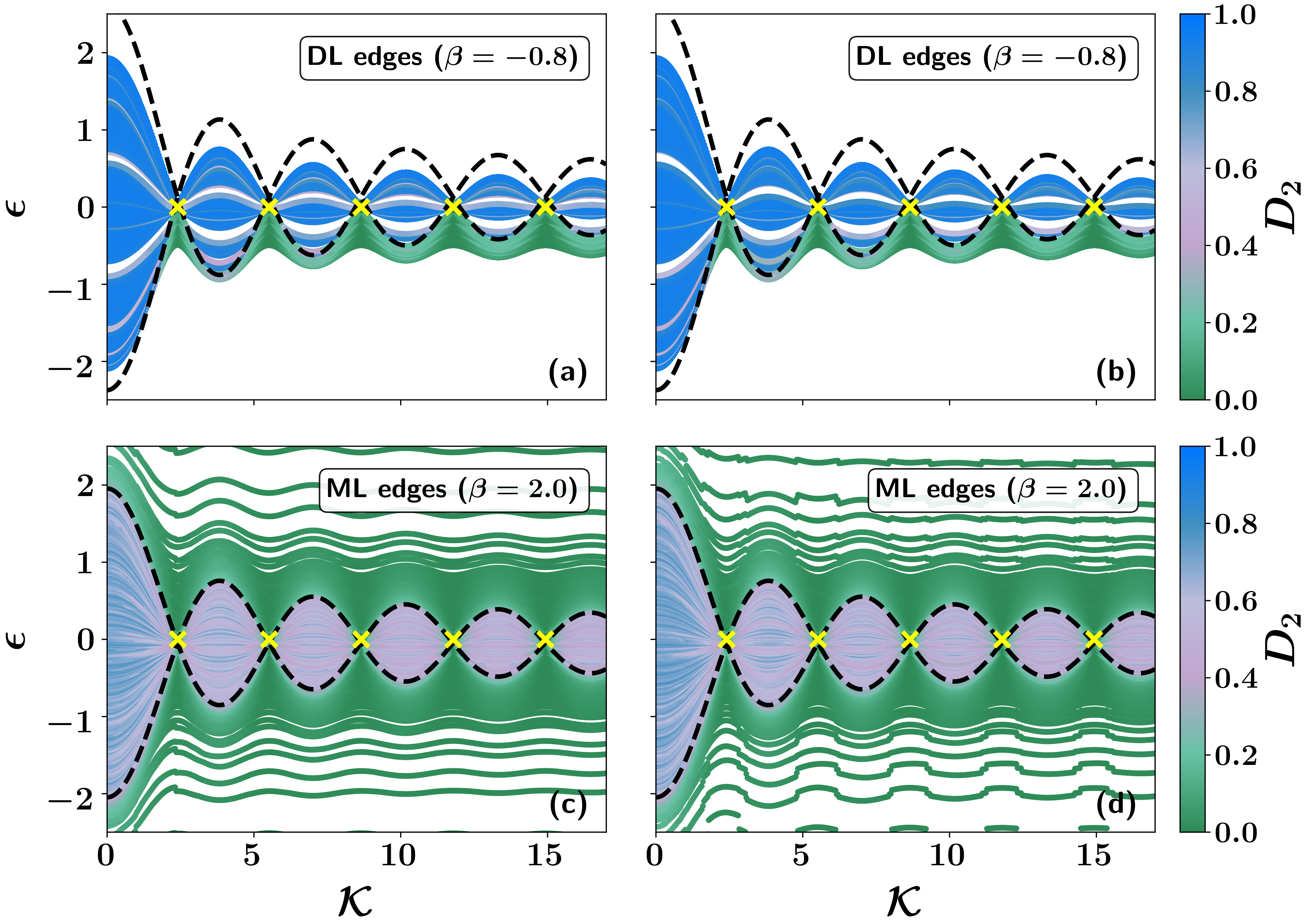}
\caption{
Fractal dimension $D_2$ of all the Floquet eigenstates of the driven GPD model for $\omega=10$ shown as a function of quasienergy $\epsilon$ and $\mathcal{K}$. 
(a),(c) For periodic boundary conditions (PBC) with $\beta=-0.8$ and $\beta=2.0$, respectively. 
(b),(d) For open boundary conditions (OBC) with $\beta=-0.8$ and $\beta=2.0$, respectively. 
Black dashed lines denote the analytically obtained mobility edges in Eq.~\eqref{Floquet_mobility_edges}. 
At the zeros of $J_0(\mathcal{K})$ (marked by yellow crosses), drive-induced localization leads to a fully localized spectrum. 
Results are shown for a system size $L = 987$, $\lambda=0.05$, and averaged over 10 realizations of the phase $\phi$.
}
\label{boundary_conditions}
\end{figure}

Here, we compare the mobility edges obtained under periodic boundary conditions (PBC) and open boundary conditions (OBC). We have plotted the $D_2$ of the Floquet eigenstates for both PBC and OBC in Fig.~\ref{boundary_conditions} in the two different regimes $\beta=-0.8$ and $\beta=2.0$. Our analysis shows that the bulk localization properties remain robust to the choice of boundary conditions. 
In particular, the analytically predicted mobility edges consistently separate the localized, multifractal, and extended regions in both cases.
We observe only minor quantitative differences between the two boundary conditions. 
For $\beta = 2.0$, the quasienergy spectrum under OBC exhibits slightly enhanced folding compared to the PBC case. 
However, this does not alter the nature of the eigenstates at the corresponding energies, and the localization properties remain unchanged. 
We have also verified this behavior using dynamical calculations with periodic boundary conditions (PBC), which agree with the results for open boundary conditions (OBC) presented in Sec.~\ref{sec:Dynamics}.
\section{Effect of quasiperiodic strength on mobility edge structures}
\label{app:lambda_drive}
\begin{figure}[t]
\includegraphics[width=\columnwidth]{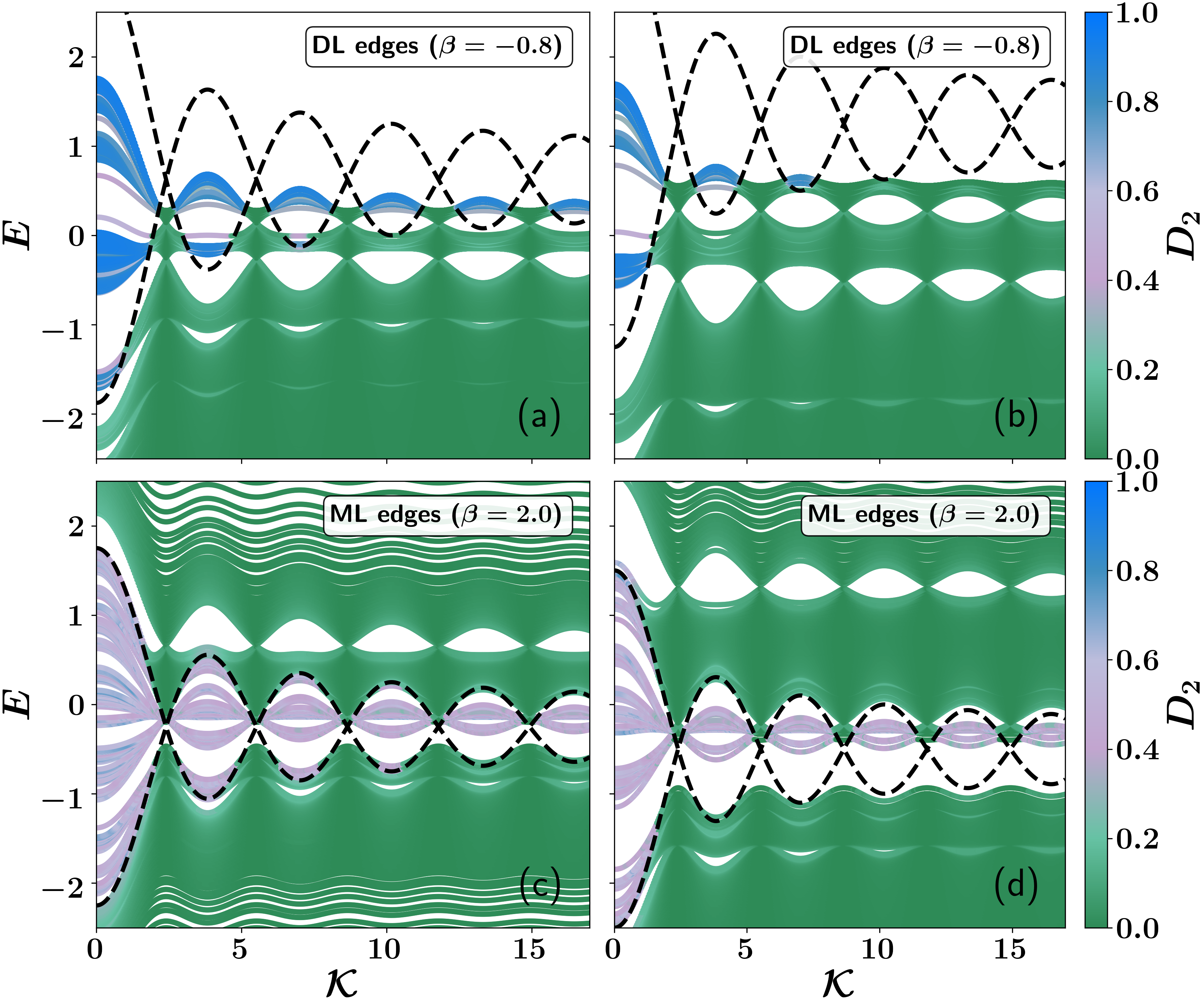}
\caption{
Fractal dimension $D_2$ of all the eigenstates of the effective Hamiltonian in Eq.~\eqref{floquet_effective} as a function of eigenenergy $E$ and $\mathcal{K}$. 
(a),(c) For $\lambda=0.25$ with $\beta=-0.8$ and $\beta=2.0$, respectively. 
(b),(d) For $\lambda=0.5$ with $\beta=-0.8$ and $\beta=2.0$, respectively. 
Results are shown for a system size $L = 987$, and averaged over 10 realizations of the phase $\phi$.
}
\label{different_lambda} 
\end{figure}
In this appendix, we analyze how the quasiperiodic potential strength $\lambda$ and periodic driving jointly influence the mobility edge structure of the system. Our discussion is based on the high-frequency effective Hamiltonian in Eq.~\eqref{floquet_effective}, where the hopping strength is renormalized as $J \to J_{\mathrm{eff}} = J J_0(\mathcal{K})$.
From the analytical results presented in Sec.~\ref{sec:floquet_ME}, the mobility edges are determined by Eq~\eqref{Floquet_mobility_edges}.
This shows that the position of the mobility edges depends explicitly on both $\lambda$ and the effective hopping $J_{\mathrm{eff}}$. 
In Fig.\ref{different_lambda}, we consider two representative values of the quasiperiodic strength, $\lambda = 0.25$ and $\lambda = 0.5$. 
Increasing $\lambda$ enhances the potential strength relative to the hopping energy set by $J_{\mathrm{eff}}$, thereby expanding the localized region in the energy spectrum. 
For a fixed driving parameter $\mathcal{K}$, the dimensionless parameter $P$ shifts systematically with $\lambda$, resulting in a displacement of the mobility edges in energy. 
Larger $\lambda$ increases $|P|$, causing a greater portion of the spectrum to satisfy $P^2 > 4\beta^2$, and hence to become localized.
Periodic driving enters through the renormalization of the hopping amplitude via $J_0(\mathcal{K})$. 
As $\mathcal{K}$ is varied, the modulation of $J_{\mathrm{eff}}$ leads to a continuous shift of the mobility edges. 
In particular, suppression of $J_0(\mathcal{K})$ reduces the effective hopping and promotes localization, while near its zeros, the system approaches a strong localization regime with $J_{\mathrm{eff}} \to 0$, where most states become localized. 
Thus, the interplay between $\lambda$ and $J_{\mathrm{eff}}$ governs the localization properties, with the drive providing a tunable control over the transition.

To further characterize the interplay between quasiperiodicity and driving, we construct the phase diagram in the $(\lambda, \mathcal{K})$ plane using the fractal dimension averaged over all eigenstates, denoted by $\langle D_2 \rangle$. 
Increasing $\lambda$ drives the system towards localization, while increasing $\mathcal{K}$ also enhances localization through the suppression of the effective hopping. 
At special points corresponding to $J_0(\mathcal{K})=0$, all states become localized. 
Away from these points, $\langle D_2 \rangle$ reflects an average over coexisting phases. In Fig.~\ref{interplay} (a), the value of $\langle D_2 \rangle$ reflects a mixture of delocalized and localized states, whereas in Fig.~\ref{interplay} (b), it reflects a mixture of localized and multifractal states. Thus the transport and localization at each point will be goverened by $\langle D_2 \rangle$ at that point, which changes due to change in the fraction of localized, delocalized and multifractal states. We note that these results are obtained using the leading-order effective Hamiltonian and are therefore expected to be reliable primarily in the high-frequency regime.  
\section{Return probability}
\label{app:return_prob}
\begin{figure}
    \centering
    \includegraphics[width=0.48\textwidth]{12.png}

\caption{
Average fractal dimension $\langle D_2 \rangle$ of all eigenstates of the effective Hamiltonian in Eq.~\eqref{floquet_effective}, plotted as a function of quasiperiodic potential strength $\lambda$ and $\mathcal{K}$ for (a) $\beta=-0.8$, and (b) $\beta=2.0$. 
Results are shown for a system size $L = 987$, averaged over 10 realizations of the phase $\phi$.
}
    
    \label{interplay}
\end{figure}

\begin{figure}[t]
\centering
\includegraphics[width=\columnwidth]{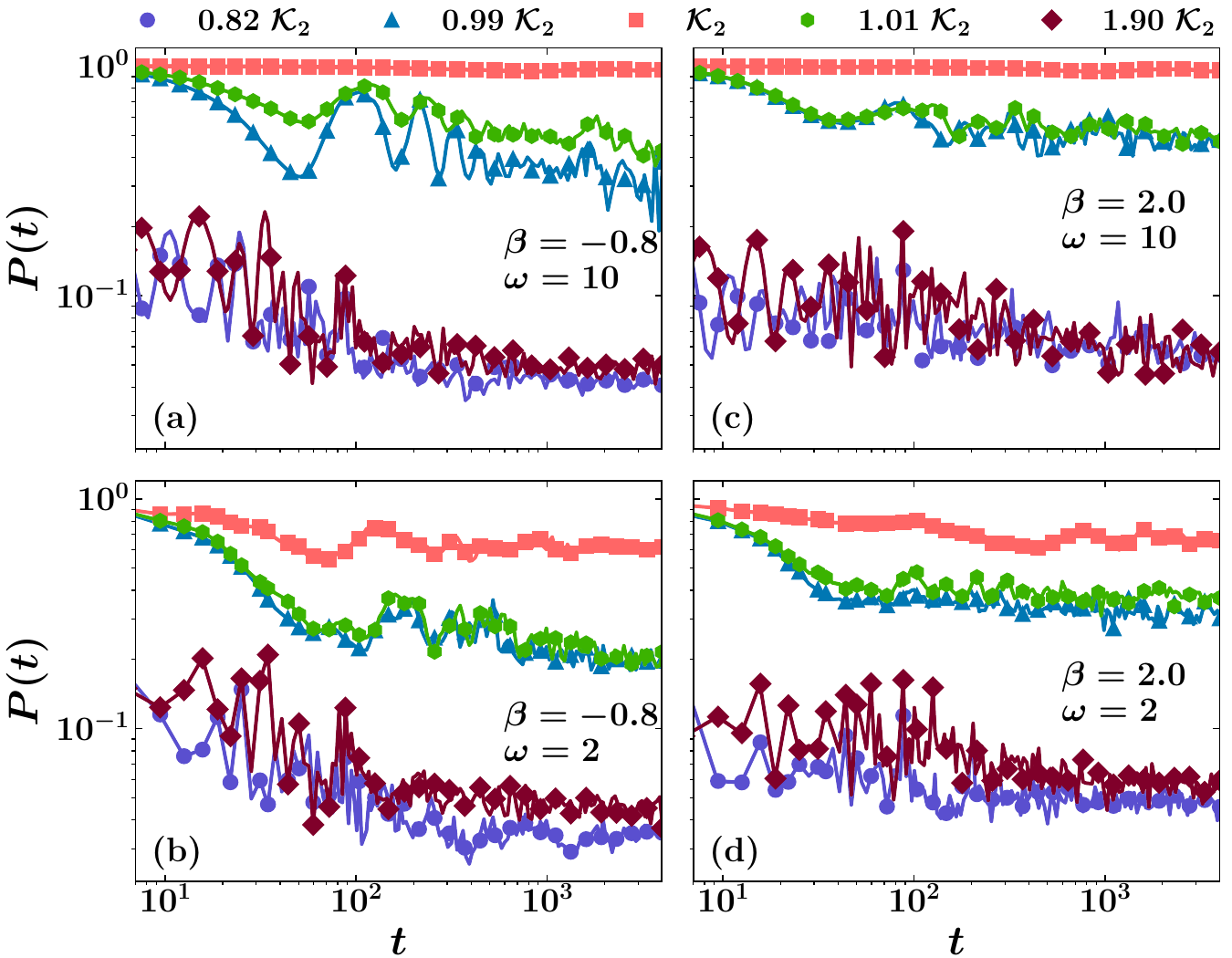}
\caption{
The return probability $P(t)$ with time $t$ for a wave packet initially localized at the center of a chain. Rows correspond to different values of the driving frequency $\omega$, while columns correspond to different values of $\beta$. Different values of $\mathcal{K}$ are represented by distinct markers and colors, with $\mathcal{K}_2 \approx 5.52$ denoting the second zero of the Bessel function $J_0(\mathcal{K})$. Results are shown for a system size $L = 987$, $\lambda=0.05$, and averaged over 100 realizations of the phase $\phi$.}
\label{fig:RET_PROB}
\end{figure}

In this appendix, we further characterize the dynamical signatures of localization and mobility edges by analyzing the return probability of an initially localized wave packet. We initialize the system in a single particle state defined in Eq.~\eqref{initial_state} and evolve it under the Floquet operator. The return probability is defined as
\begin{equation}
P(t) = |\langle \psi(0) | \psi(t) \rangle|^2 ,
\end{equation}
which quantifies the memory of the initial state. In extended regimes $P(t)$ decays rapidly, while a finite value close to $1$ at long-times signals localization. Fig.~\ref{fig:RET_PROB} shows the time evolution of $P(t)$ for both DL and ML edges at two representative driving frequencies.

For the DL edge corresponding to $\beta=-0.8$ [Fig.~\ref{fig:RET_PROB}(a) and (b)], away from the drive-induced localization point $\mathcal{K}_2$, $P(t)$ exhibits a rapid decay from unity, followed by oscillatory behavior at intermediate times before saturating at small values for both driving frequencies, indicating strong delocalization due to the dominance of extended states. 
At $\mathcal{K}=\mathcal{K}_2$, $P(t)$ remains close to unity for $\omega=10$, demonstrating strong confinement of the wave packet. In contrast, for $\omega=2$, the saturation value is significantly reduced and exhibits stronger fluctuations, indicating enhanced transport due to higher-order processes.
The same behavior is observed for the regime associated with ML edges with $\beta=2.0$ [Fig~\ref{fig:RET_PROB}(c)-(d)].
\section{Derivation of Higher-Order Corrections in the Floquet--Magnus Expansion}
\label{app:low_freq}

A comparison between the two driving frequencies shows that lowering the driving frequency enhances localization within the multifractal band. Also, for both the DL and ML edge cases, the drive-induced localization at the zeros of the Bessel function is not perfect and does not suppress transport entirely, as we have seen from Sec.~\ref{sec:Dynamics}. Moreover, the suppression becomes weaker as we lower the driving frequency.  To capture these low-frequency effects, we derive the next higher-order contribution to the Floquet Hamiltonian.

We consider spinless fermions with nearest-neighbor hopping in a quasiperiodic potential, where we define the bond operators as
\begin{equation}
\hat{T}_j = \hat{c}_j^\dagger \hat{c}_{j+1}, 
\qquad 
\hat{T}_j^\dagger = \hat{c}_{j+1}^\dagger \hat{c}_j,
\end{equation}
and the number operator as
\begin{equation}
\hat{n}_j = \hat{c}_j^\dagger \hat{c}_j.
\end{equation}
The rotating frame Hamiltonian in Eq.~\eqref{rotating_frame_Hamiltonian} is thus given by
\begin{equation}
\hat{H}_R(t)
=
\sum_j
\Big[
J e^{iF(t)} \hat{T}_j
+
J e^{-iF(t)} \hat{T}_j^\dagger
\Big]
+
\sum_j V_j \hat{n}_j .
\end{equation}
Now using the Jacobi--Anger expansion
\begin{equation}
e^{i x \sin\theta} = \sum_{m=-\infty}^{\infty} J_m\!\left(\frac{K}{\omega}\right) e^{i m \theta},
\end{equation}
the Hamiltonian $\hat{H}_R(t)$ can be written as a Fourier sum,
\begin{equation}
\hat{H}_R(t) = \sum_m \hat{H}_m e^{i m \omega t}.
\end{equation}
Explicitly,
\begin{equation}
\begin{aligned}
\hat{H}_R(t)
&=
\sum_{j,m}
\Big[
J J_m\!\left(\frac{K}{\omega}\right) e^{i m \omega t} \hat{T}_j
+
J J_m\!\left(\frac{K}{\omega}\right) e^{-i m \omega t} \hat{T}_j^\dagger
\Big] \\
&\quad
+
\sum_j V_j \hat{n}_j.
\end{aligned}
\end{equation}
Thus, the Fourier components of the driven Hamiltonian are
\begin{equation}
\label{fourier_components}
\hat{H}_m = J J_m(\mathcal{K}) \sum_l \left( \hat{T}_l + (-1)^m \hat{T}_l^\dagger \right),
\end{equation}
\begin{equation}
\hat{H}_{-m} = J J_m(\mathcal{K}) \sum_l \left( \hat{T}_l^\dagger + (-1)^m \hat{T}_l \right),
\end{equation}
where $\mathcal{K} = K/\omega$. To obtain these expressions, we have used the identity $J_{-m} = (-1)^m J_m$.
The time evolution operator over one driving period, obtained from the rotating-frame Hamiltonian, can be expressed as
\begin{equation}
\hat{U}(T) = \exp\left(-i \hat{H}_F T \right),
\end{equation}
which defines the effective Floquet Hamiltonian $\hat{H}_F$.
The Floquet Hamiltonian admits the expansion~\cite{Eckardt_2015,Sen_2021}
\begin{equation}
\hat{H}_F = \hat{H}_F^{(0)} + \hat{H}_F^{(1)} + \hat{H}_F^{(2)} + \mathcal{O}(\omega^{-3}) + \cdots
\end{equation}
with
\begin{align}
\hat{H}_F^{(0)} &= \hat{H}_0, \\
\hat{H}_F^{(1)} &= \sum_{m\neq 0} \frac{[\hat{H}_{-m},\hat{H}_m]}{2m\omega}, \\
\hat{H}_F^{(2)} &= \sum_{m\neq 0} \frac{[\hat{H}_{-m},[\hat{H}_0,\hat{H}_m]]}{2(m\omega)^2}
+ \sum_{\substack{m\neq 0 \\ n\neq 0,m}} \frac{[\hat{H}_{-m},[\hat{H}_{m-n},\hat{H}_n]]}{3mn\omega^2}.
\end{align}
Thus, the zeroth-order term is given by
\begin{equation}
\hat{H}_F^{(0)} = \hat{H}_0 = J J_0\!\left(\mathcal{K}\right)\sum_j (\hat{T}_j + \hat{T}_j^\dagger) + \sum_j V_j \hat{n}_j,
\end{equation}
which corresponds to the Hamiltonian in Eq.~\eqref{floquet_effective}.
Next, we turn to calculate the higher-order terms. To this end, we make use of the fermionic anticommutation relations,
\begin{equation}
\{\hat{c}_i, \hat{c}_j^\dagger\} = \delta_{ij}, \qquad
\{\hat{c}_i, \hat{c}_j\} = 0, \qquad
\{\hat{c}_i^\dagger, \hat{c}_j^\dagger\} = 0.
\end{equation}
From these relations, one obtains the following commutators:
\begin{equation}
\label{commutators}
\begin{aligned}
[\hat{T}_j, \hat{T}_l] &=
\hat{c}_j^\dagger \hat{c}_{j+2}\,\delta_{l,j+1}
- \hat{c}_{j-1}^\dagger \hat{c}_{j+1}\,\delta_{l,j-1}, \\
[\hat{T}_j^\dagger, \hat{T}_l^\dagger] &= 
- \hat{c}_{j+2}^\dagger \hat{c}_j\,\delta_{l,j+1}
+ \hat{c}_{j+1}^\dagger \hat{c}_{j-1}\,\delta_{l,j-1}, \\
[\hat{T}_j, \hat{T}_l^\dagger] &= (\hat{n}_j - \hat{n}_{j+1})\delta_{l,j}, \\
[\hat{T}_j^\dagger, \hat{T}_l] &= (\hat{n}_{j+1} - \hat{n}_j)\delta_{l,j}, \\
[\hat{n}_j,\hat{T}_l] &= (\delta_{j,l}-\delta_{j,l+1})\hat{T}_l, \\
[\hat{n}_j,\hat{T}_l^\dagger] &= (\delta_{j,l+1}-\delta_{j,l})\hat{T}_l^\dagger.
\end{aligned}
\end{equation}
This set of commutators will be used extensively in the calculation of the higher order contributions to the Floquet Hamiltonian.
Now consider the first-order Floquet term
\begin{equation}
\hat{H}_F^{(1)} = \sum_{m\neq 0} \frac{[\hat{H}_{-m},\hat{H}_m]}{2m\omega}.
\end{equation}
Substituting the Fourier components, the commutator $[\hat{H}_{-m},\hat{H}_m]$ takes the form
\begin{equation}
\begin{aligned}
[\hat{H}_{-m},\hat{H}_m] &= J^2 J_m^2\!\left(\mathcal{K}\right)
\sum_{j,l}
\Big(
[\hat{T}_j^\dagger,\hat{T}_l]
+(-1)^m [\hat{T}_j^\dagger,\hat{T}_l^\dagger] \\
&\qquad
+(-1)^m [\hat{T}_j,\hat{T}_l]
+[\hat{T}_j,\hat{T}_l^\dagger]
\Big).
\end{aligned}
\end{equation}
Using the commutators in Eq.~\eqref{commutators}, this simplifies to
\begin{equation}
\begin{aligned}
[\hat{H}_{-m},\hat{H}_m]
&=
J^2 J_m^2\!\left(\mathcal{K}\right)(-1)^m
\sum_j
\Big(
- \hat{c}_{j+1}^\dagger \hat{c}_{j-1}
+ \hat{c}_{j+2}^\dagger \hat{c}_{j} \\
&\qquad
+ \hat{c}_{j-1}^\dagger \hat{c}_{j+1}
- \hat{c}_{j}^\dagger \hat{c}_{j+2}
\Big)
\end{aligned}
\end{equation}
Now, replacing the dummy index $j \to j+1$ in the first and the third term inside the summation, we obtain an exact cancellation of all operator terms for periodic boundary conditions. For open boundary conditions, only edge terms survive, which become insignificant for probing bulk localization physics. In the thermodynamic limit, these contributions can therefore be neglected.
Furthermore, even without invoking operator-level cancellation, the Floquet sum satisfies
\begin{equation}
\sum_{m\neq 0} \frac{J_m^2(\mathcal{K})}{m}
= \sum_{m>0} \left( \frac{J_m^2(\mathcal{K})}{m} + \frac{J_{-m}^2(\mathcal{K})}{-m} \right) = 0,
\end{equation}
since $J_{-m}^2(\mathcal{K}) = J_m^2(\mathcal{K})$. Therefore,
\begin{equation}
\hat{H}_F^{(1)} = 0.
\end{equation}

Since the first-order term vanishes, the leading correction arises at second order. We therefore compute the second-order Floquet contribution,
\begin{equation}
\hat{H}_F^{(2)} =
\sum_{m\neq 0} \frac{[\hat{H}_{-m},[\hat{H}_0,\hat{H}_m]]}{2(m\omega)^2}
+ \sum_{m\neq 0}\sum_{n\neq 0,m} \frac{[\hat{H}_{-m},[\hat{H}_{m-n},\hat{H}_n]]}{3mn\omega^2}.
\end{equation}
The second term in the above equation vanishes due to operator-level cancellation, analogous to the first-order case. Hence, the remaining contribution arises solely from the first term.
We compute
\begin{equation}
[\hat{H}_0,\hat{H}_m]
=
J J_m\!\left(\mathcal{K}\right)
\sum_{j,l}
\Big(
V_j[\hat{n}_j,\hat{T}_l]
+ (-1)^m V_j[\hat{n}_j,\hat{T}_l^\dagger]
\Big).
\end{equation}
Evaluating the commutators suing Eq.~\eqref{commutators}, this simplifies to
\begin{equation}
[\hat{H}_0,\hat{H}_m]
=
J J_m\!\left(\mathcal{K}\right)
\sum_l (V_l - V_{l+1})
\left(\hat{T}_l - (-1)^m \hat{T}_l^\dagger \right).
\end{equation}
The next commutator to evaluate is
\begin{equation}
\begin{aligned}
[\hat{H}_{-m},[\hat{H}_0,\hat{H}_m]]
&=
J^2 J_m^2\!\left(\mathcal{K}\right)
\Bigg[
\sum_j (\hat{T}_j^\dagger + (-1)^m \hat{T}_j), \\
&\qquad
\sum_l (V_l - V_{l+1})(\hat{T}_l - (-1)^m \hat{T}_l^\dagger)
\Bigg]
\end{aligned}.
\end{equation}
Carrying out the algebra (term-by-term using the commutation relations), this gives
\begin{widetext}
\begin{equation*}
\begin{aligned}
[\hat{H}_{-m},[\hat{H}_0,\hat{H}_m]] 
&= J^2 J_m^2\!\left(\mathcal{K}\right)
\Bigg[
\sum_j 2(V_j - V_{j+1})(\hat{n}_{j+1} - \hat{n}_j)
+ (-1)^m \sum_j (V_{j} - V_{j-1})
(\hat{c}_{j+1}^\dagger \hat{c}_{j-1} + \hat{c}_{j-1}^\dagger \hat{c}_{j+1})
\\
&\qquad
+ (-1)^m \sum_j (V_{j+1} - V_{j+2})
(\hat{c}_{j+2}^\dagger \hat{c}_{j} + \hat{c}_{j}^\dagger \hat{c}_{j+2})
\Bigg].
\end{aligned}
\end{equation*}

Rewriting by shifting indices,

\begin{equation*}
\begin{aligned}
[\hat{H}_{-m},[\hat{H}_0,\hat{H}_m]] 
&= J^2 J_m^2\!\left(\mathcal{K}\right)
\Bigg[
\sum_j 2(V_{j+1} + V_{j-1} - 2V_j)\hat{n}_j
+ (-1)^m \sum_j (2V_{j+1} - V_j - V_{j+2})
(\hat{c}_j^\dagger \hat{c}_{j+2} + \hat{c}_{j+2}^\dagger \hat{c}_j)
\Bigg].
\end{aligned}
\end{equation*}
Thus, the second-order correction is given by
\begin{equation}
\begin{aligned}
\hat{H}_F^{(2)} &=
\sum_{m\neq 0}
\frac{
J^2 J_m^2\!\left(\mathcal{K}\right)
}{
2(m\omega)^2
}
\Bigg[
\sum_j 2(V_{j+1}+V_{j-1}-2V_j)\hat{n}_j
+ (-1)^m \sum_j (2V_{j+1}-V_j-V_{j+2})
(\hat{c}_j^\dagger \hat{c}_{j+2} + \hat{c}_{j+2}^\dagger \hat{c}_j)
\Bigg].
\end{aligned}
\end{equation}
\end{widetext}
The second-order term introduces two competing physical effects. The first contribution modifies the onsite potential, providing qualitative insight into why the multifractal band exhibits enhanced localization at lower frequencies. The second contribution generates emergent next-nearest-neighbor hoppings, which tend to facilitate transport, an effect particularly pronounced near the drive-induced localization points. This helps explain why perfect dynamical localization is not strictly achieved at finite frequencies, even at the zeros of the Bessel function. While exact numerical evolution at lower frequencies ($\omega=2$) incorporates all-order Floquet processes, $\hat{H}_F^{(2)}$ serves as the leading non-zero correction that illustrates these essential physical mechanisms.

\section{Stability of Growth Exponents Against System Sizes and Fitting Windows}
\label{app:exponent_stability}

To quantify the reliability of the transport exponents ($\gamma$ for the root-mean-squared displacement $X(t) \sim t^{\gamma}$ and $\nu$ for the entanglement entropy $S(t) \sim t^{\nu}$), we evaluate their stability under variations in system size and choice of fitting windows. In all cases, power-law exponents are extracted by performing linear fits to $\log X(t)$ and $\log S(t)$ versus $\log t$ in late-time scaling regimes chosen strictly after the decay of initial transient behavior and well before the onset of finite-size saturation.

\begin{figure}[t]
\centering
\includegraphics[width=\columnwidth]{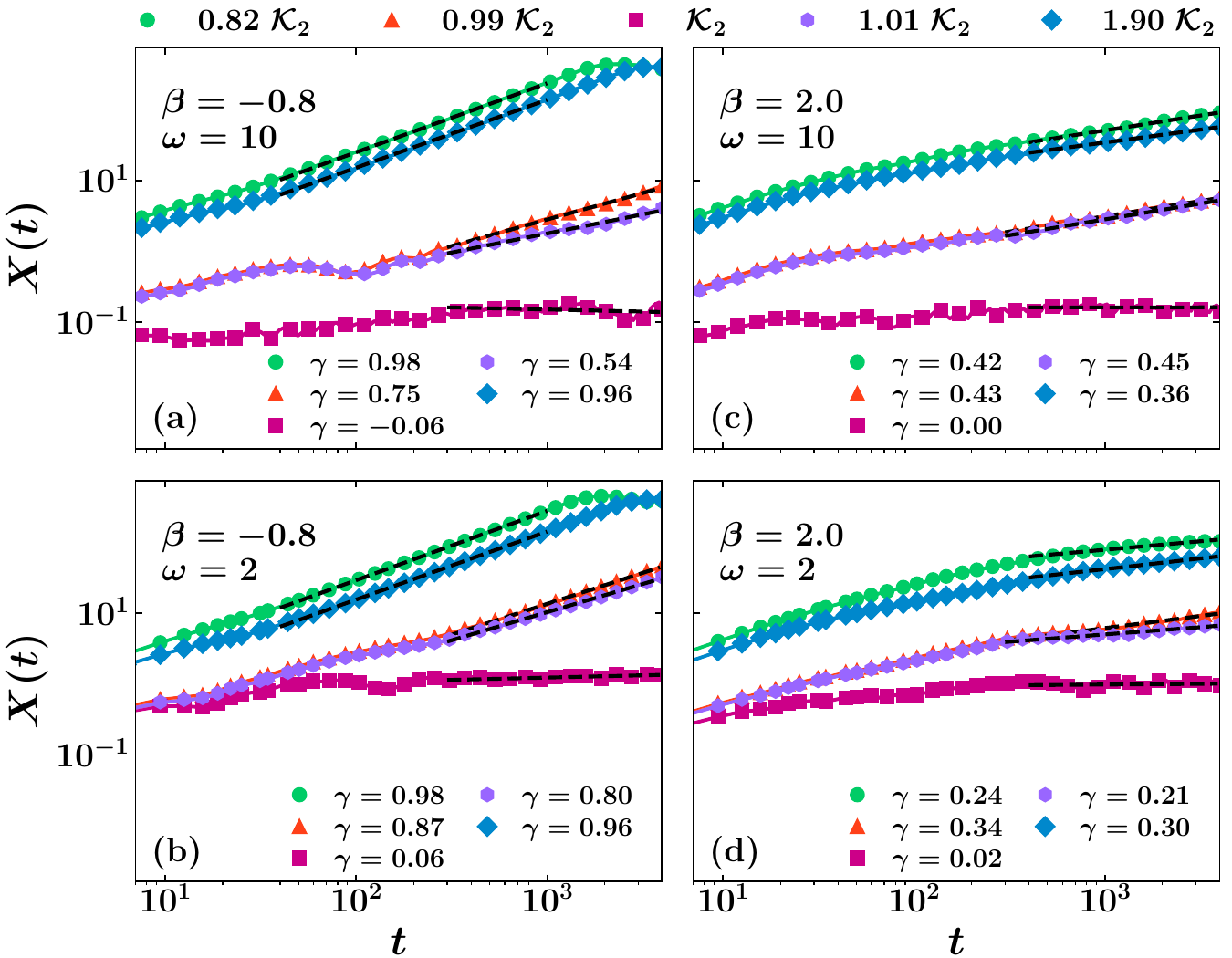}
\caption{
Dynamics of the root-mean-squared displacement $X(t)$ for an increased system size $L = 1597$. Rows correspond to $\omega = 10$ and $\omega = 2$, while columns correspond to $\beta = -0.8$ and $\beta = 2.0$. Distinct markers and colors represent different values of $\mathcal{K}$. 
 lines denote power-law fits $X(t) \sim t^{\gamma}$ evaluated over the same time windows as used in Fig.~\ref{fig:RMSD} for $L = 987$. The extracted exponents $\gamma$ differ by at most $\approx 0.04$ from the $L = 987$ baseline, demonstrating system-size invariance. All results are for $\lambda = 0.05$ and averaged over 100 realizations of the phase $\phi$.}
\label{fig:RMSD_L1597}
\end{figure}
\begin{figure}[t]
\centering
\includegraphics[width=\columnwidth]{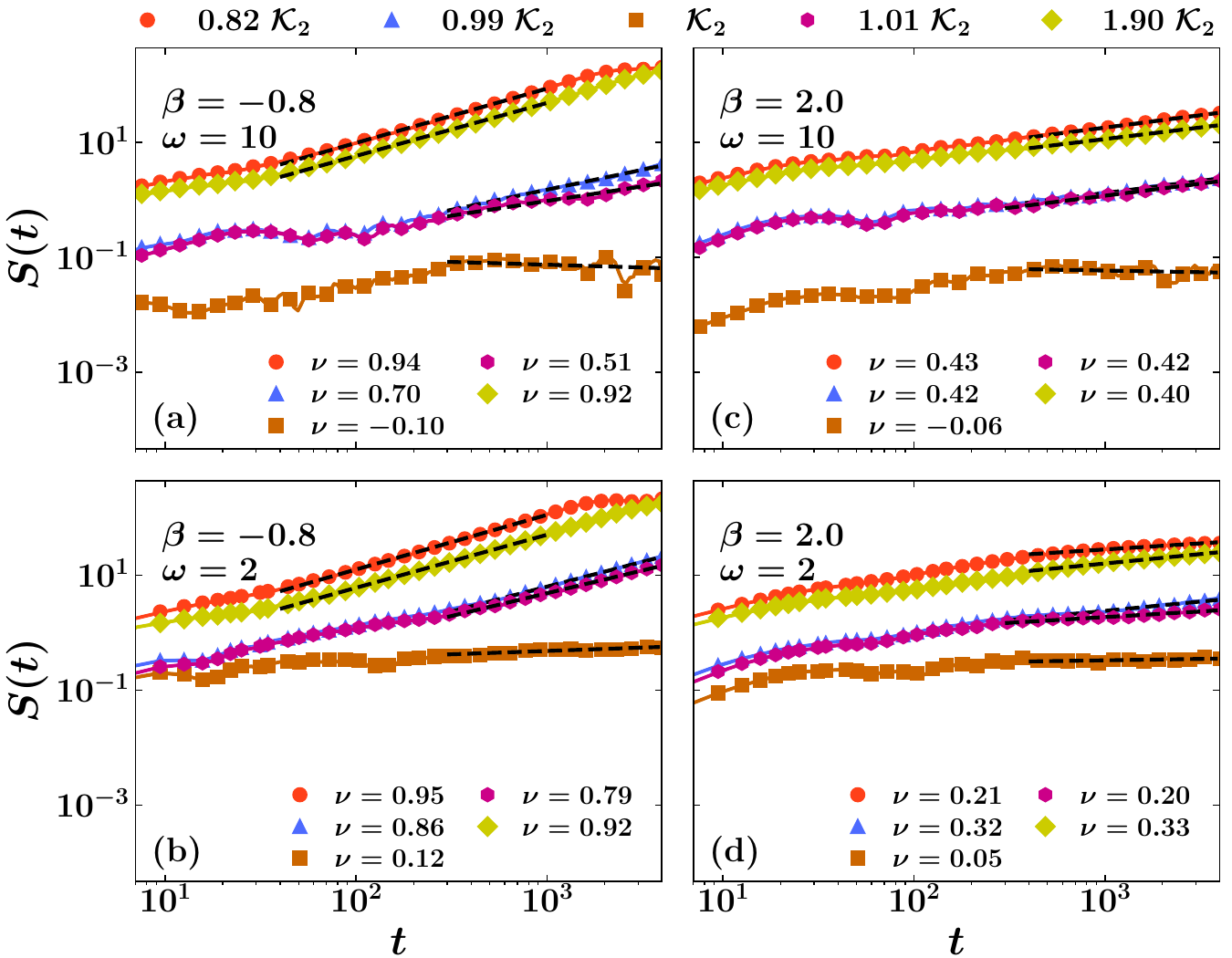}
\caption{
Dynamics of the half-chain entanglement entropy $S(t)$ for an increased system size $L = 1597$. Panel layouts, color schemes, and marker choices mirror those in Fig.~\ref{fig:RMSD_L1597}. Black dashed lines denote power-law fits $S(t) \sim t^{\nu}$ evaluated over the same time windows as used in Fig.~\ref{fig:EE}. The extracted exponents $\nu$ vary by at most $\approx 0.05$ compared to $L = 987$, confirming that entanglement growth rates are robust with respect to change in system size. All results are for $\lambda = 0.05$ and averaged over 100 realizations of the phase $\phi$.}
\label{fig:EE_L1597}
\end{figure}

\begin{figure*}[t]
\centering
\begin{minipage}[t]{0.48\textwidth}
\centering
\includegraphics[width=\linewidth]{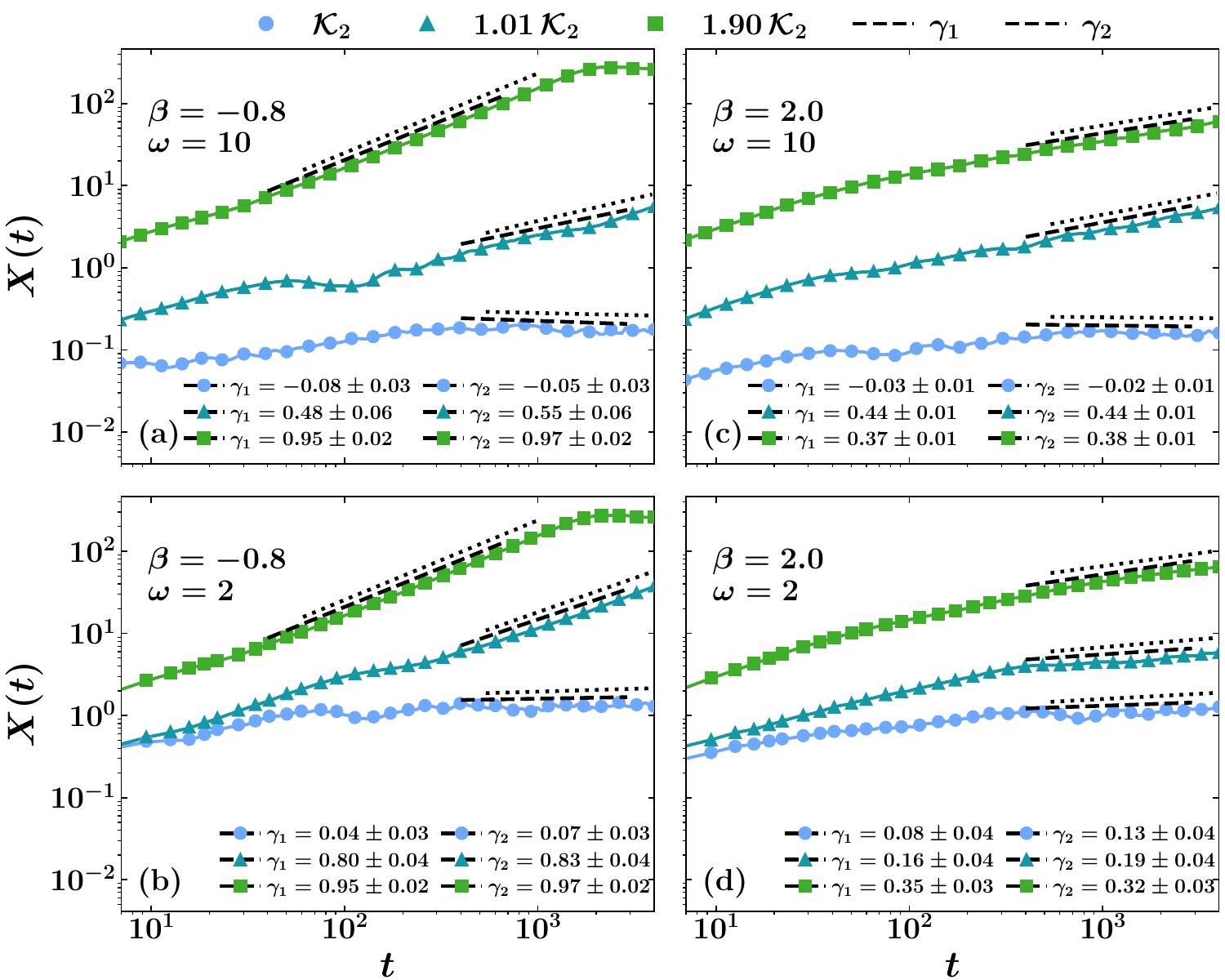}
\caption{
Sensitivity of the RMSD growth exponent $\gamma$ to variations in the fitting window for system size $L = 987$. Fits are performed over two overlapping late-time windows: Window 1 ($\gamma_1$, black dashed lines) and Window 2 ($\gamma_2$, black dotted lines). Extracted values and inter-window uncertainties $\gamma_1 \pm \Delta\gamma$ and $\gamma_2 \pm \Delta\gamma$ are displayed in the panel legends. Discrepancies stay within $0.01\text{--}0.04$ across almost all panels, demonstrating stability against window selection.
}
\label{fig:RMSD_windows}
\end{minipage}%
\hfill
\begin{minipage}[t]{0.48\textwidth}
\centering
\includegraphics[width=\linewidth]{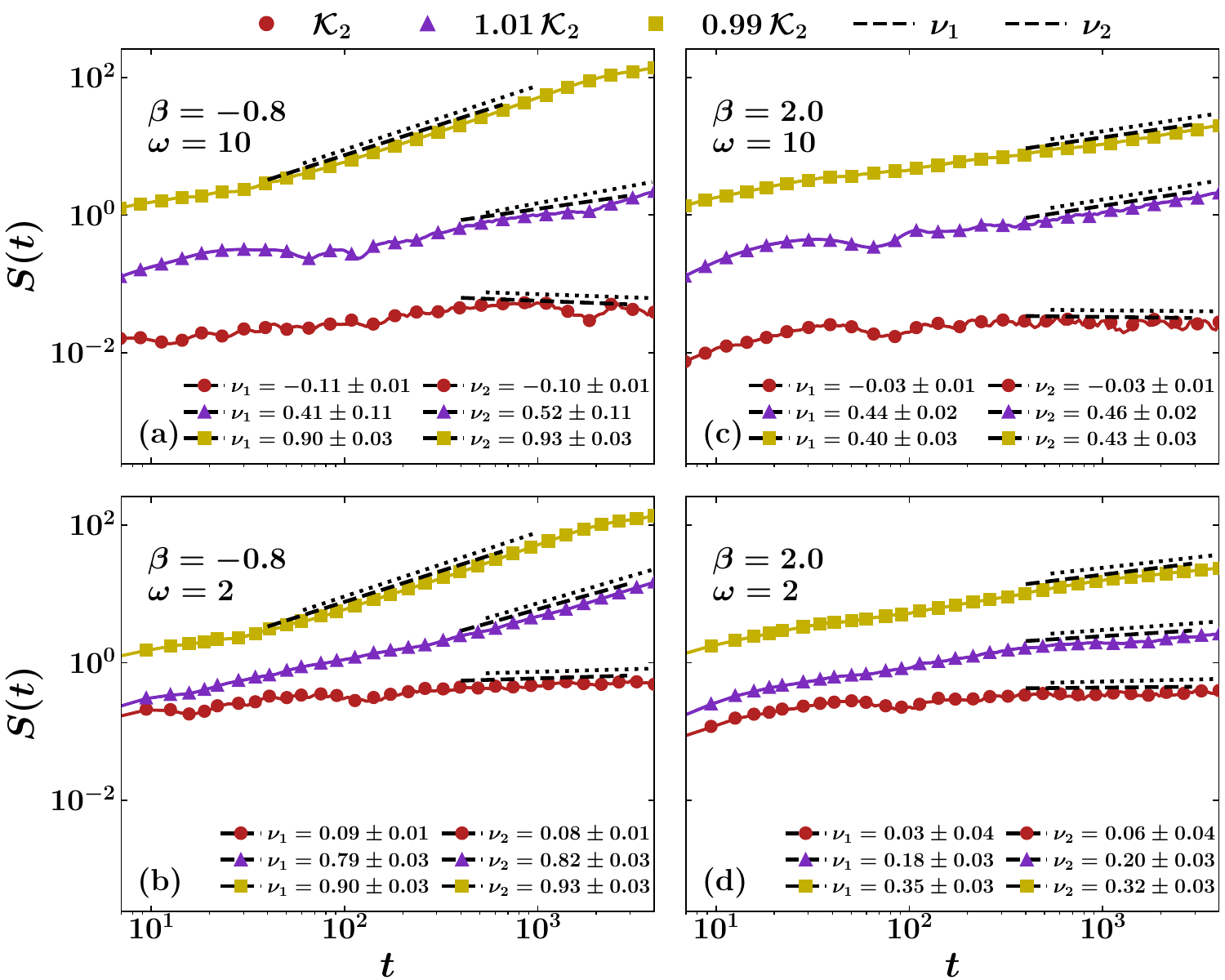}
\caption{
Sensitivity of the entanglement entropy growth exponent $\nu$ to variations in the fitting window for system size $L = 987$. Fits are performed over two overlapping late-time windows: Window 1 ($\gamma_1$, black dashed lines) and Window 2 ($\gamma_2$, black dotted lines). Legend values report $\nu_1 \pm \Delta\nu$ and $\nu_2 \pm \Delta\nu$. Inter-window exponent variations remain bounded within $0.01\text{--}0.04$, establishing the quantitative stability of the entropy growth rates.
}
\label{fig:EE_windows}
\end{minipage}
\end{figure*}
\subsection{Robustness Against System Size Variations}
To confirm that the extracted exponents represent asymptotic thermodynamic-limit transport rather than finite-size artifacts, we perform identical dynamical simulations for an expanded system size of $L = 1597$ (compared to $L = 987$ reported in the main text). Fitting is conducted using the exact same time windows adopted for $L = 987$.
Comparing the extracted exponents across system sizes reveals high quantitative stability. The maximum deviation in the RMSD growth exponent $\gamma$ across all driving parameters $\mathcal{K}$, $\omega$, and quasiperiodic potential parameter $\beta$ does not exceed $\Delta \gamma \approx 0.04$. Similarly, the entanglement entropy exponent $\nu$ shifts by at most $\Delta \nu \approx 0.05$. Crucially, these minor numeric shifts leave the physical transport classifications intact: superdiffusive, subdiffusive, and near-ballistic regimes remain strictly robust across system sizes. The dynamics for $L = 1597$ are displayed in Figs.~\ref{fig:RMSD_L1597} and \ref{fig:EE_L1597}.

\subsection{Robustness Against Fitting Window Variations}
To address potential sensitivity to the choice of time intervals, we perform a dual-window stability analysis for three representative driving strengths: $\mathcal{K} = \mathcal{K}_2$, $\mathcal{K} = 1.01\mathcal{K}_2$, and $\mathcal{K} = 1.90\mathcal{K}_2$. For each case, power-law exponents are extracted over two adjacent late-time windows. Window 1 (black dashed lines) is selected immediately after initial transient dynamics subside. Window 2 (black dotted lines) extends deeper into the late-time regime prior to finite-size saturation, serving to test exponent stability over longer timescales. The extracted values ($\gamma_1, \nu_1$ for Window 1 and $\gamma_2, \nu_2$ for Window 2) along with their inter-window variations $\Delta \gamma = |\gamma_1 - \gamma_2|$ and $\Delta \nu = |\nu_1 - \nu_2|$ are displayed directly in the panel legends of Figs.~\ref{fig:RMSD_windows} and \ref{fig:EE_windows}.
Across nearly all parameter sets, the extracted exponents exhibit high stability against window selection, with variations remaining small ($0.01 \le \Delta \gamma, \Delta \nu \le 0.06$).
This confirms that the long-time transport classifications remain physically sound and robust.

\bibliography{refs}

\end{document}